# Optimization of Bound Disjunctive Queries with Constraints*

Gianluigi Greco, Sergio Greco, Irina Trubitsyna and Ester Zumpano

*DEIS*
*Università della Calabria*
*87030 Rende, Italy*
*e-mail: {ggreco,greco,irina,zumpano}@si.deis.unical.it*

## Abstract

This paper presents a technique for the optimization of bound queries over disjunctive deductive databases with constraints. The proposed approach is an extension of the well-known Magic-Set technique and is well-suited for being integrated in current bottom-up (stable) model inference engines. More specifically, it is based on the exploitation of binding propagation techniques which reduce the size of the data relevant to answer the query and, consequently, reduces both the complexity of computing a single model and the number of models to be considered. The motivation of this work stems from the observation that traditional binding propagation optimization techniques for bottom-up model generator systems, simulating the goal driven evaluation of top-down engines, are only suitable for positive (disjunctive) queries, while hard problems are expressed using unstratified negation.

The main contribution of the paper consists in the extension of a previous technique, defined for positive disjunctive queries, to queries containing both disjunctive heads and constraints (a simple and expressive form of unstratified negation). As the usual way of expressing declaratively hard problems is based on the *guess-and-check* technique, where the *guess* part is expressed by means of disjunctive rules and the *check* part is expressed by means of constraints, the technique proposed here is highly relevant for the optimization of queries expressing hard problems. The value of the technique has been proved by several experiments.

## 1 Introduction

Disjunctive Datalog programs, i.e. programs allowing clauses to have both disjunction in their heads and negation in their bodies (short. Datalog$^{\vee,\neg}$ programs), have been successfully introduced with the aim of modelling incomplete data (Lobo et al., 1992). Over the last few years, there has been a great deal of interest in studying declarative semantics for Datalog$^{\vee,\neg}$ programs. In fact, minimal model semantics (Grant & Minker, 1986) is widely accepted for programs in the absence of body negation (Datalog$^{\vee}$ programs), and is naturally extended for programs with (possibly unstratified) negation. For Datalog$^{\vee,\neg}$ programs there are several extensions of the minimal model semantics such as disjunctive stable models (Przymusinski, 1991; Gelfond & Lifschitz, 1991) minimal founded semantics (Furfaro et al., 2004), and different proposals of well-founded semantics (see, e.g., (Ross, 1990; Brass & Dix, 1999; Przymusinski, 1995; Eiter et al., 1997b; Baral et al., 1992; Wang, 2000)).

---

* A preliminary version of this paper was presented at the LPAR'02 Conference (Greco et al. 2002). The second author is also supported by ICAR-CNR.

For each of the above semantics, several algorithms and techniques for model generation have also been proposed. Model generation is often carried out through bottom-up evaluation of clauses (Minker & Rajasekar, 1990; Brass & Dix, 1994; Fernández & Minker, 1995a; Leone et al., 2002; Simons et al., 2002).

For instance, bottom-up methods have been employed to compute perfect and stable models (Fernández & Minker, 1995a; Fernández & Minker, 1995b) using ordered model trees, and to process logic programs under the minimal model semantics using a tableau calculus (Niemela, 1996).

This paper focuses on stable model semantics, according to which a disjunctive program may have several alternative models, called *answer sets*, each one corresponding to a possible view of the world — see (Niemela, 2003) for a recent overview of *answer set programming*. Disjunctive logic programs under stable model semantics are very expressive, since they capture the complexity class $\Sigma_2^P$ (Eiter et al., 1997a) and they have been used for developing several practical applications. Furthermore, in the last few years several efficient inference engines implementing stable model semantics have been developed; the DLV system (Leone et al., 2002) and the GnT system (Janhunen et al., 2000) should be recalled which implement disjunctive stable models, while for the non-disjunctive case many other engines are currently available (see, e.g., *Smodels* (Syrjanen & Niemela, 2001), DeReS (Cholewinski et al., 1996), and ASSAT (Lin & Zhao, 2002)).

Even though model generator techniques for stable model semantics are quite useful for knowledge representation and reasoning tasks, it is well-known that they are inefficient when used for refutations, i.e. query answering. In fact, they often explore a search space much larger than that required, and tend to generate answers to all the possible queries rather than to the precise query.

However, it is often the case that only a strict subset of the stable models needs to be considered and that there is no need to compute these models in their entirety to answer the query. This intuition is exploited by top-down techniques which only consider atoms necessary to answer the current query and outperform model generators used for refutation. In fact, top-down approaches systematically utilize the query to propagate the binding into the body of the rules to avoid computing all the models of the program — a brief overview of top-down (disjunctive) reasoning is presented in the next section.

In order to optimize query evaluation, while still preserving the ability to compute models (well-suited for complex reasoning tasks), several works proposed the simulation of top-down strategies, by means of suitable transformations introducing new predicates and rewriting some clauses.

Among them, the Magic-Set technique is the best-known technique. This is due to its efficiency and its generality, even though other focused methods, such as the supplementary Magic-Set and other special techniques for linear and chain queries have also been proposed (see, e.g., (Beeri & Ramakrisnhan, 1991; Greco *et al.*, 1995; Ullman, 1989a; Ramakrisnhan *et al.*, 1993)).

It should be recalled that the formal equivalence between top-down evaluation with memoing and bottom-up evaluation with Magic-Set optimization is well-known (Ullman, 1989a; Ullman, 1989b). However, this Magic-Set optimization technique is only suitable for positive Datalog queries, i.e. queries without disjunction and negation. To the best of our knowledge the first extension of the Magic-Set technique for the evaluation of disjunctive Datalog$^\vee$ queries, in which negation is not allowed, was introduced in (Greco, 1999) and extended in (Greco, 2003).

This paper further investigates optimization techniques simulating the top-down evaluation of queries, by extending the proposal presented in (Greco, 2003) from Datalog$^\vee$ programs to Datalog$^{\vee,\neg}$ programs. Actually, a syntactic restriction of Datalog$^{\vee,\neg}$ programs is considered in which unstratified negation can only be expressed in the form of constraint rules. Notice that this is not truly a restriction. In fact, constraints represent a natural way to extend database semantics, by explicitly defining properties which are supposed to be satisfied by all instances over a given database schema. Moreover, the usual way of expressing declaratively hard problems, such as NP problems and problems in the second level of the polynomial hierarchy ($\Sigma_p^2$ and $\Pi_p^2$ problems), is based on the *guess-and-check* technique where the *guess* part is expressed by means of disjunctive rules and the *check* part is expressed by means of constraints. Therefore, the technique proposed here is highly relevant for the optimization of queries expressing hard problems.

Even though constraints can be easily managed in top-down approaches, it should be pointed out they represent a serious issue when simulating top-down reasoning by means of the Magic-Set technique. In fact, all the rewriting techniques presented so far in the literature cannot be straightforwardly extended to constraint rules.

### *1.1 Contributions*

The main contributions of the paper are as follows.

- A query rewriting algorithm is defined which allows the simulation of top-down computation in bottom-up inference engines by propagating the bindings from the query-goal into the body of rules. The rewriting technique avoids the computation of complete models and useless models which are not necessary to answer the query. Essentially the technique is an adaptation of the Magic-Set technique (Bancilhon *et al.*, 1986; Beeri & Ramakrisnhan, 1991) to disjunctive Datalog programs with constraints.
- It is observed that the application of the Magic-Set technique to queries with

constraints produces a query that can be evaluated more efficiently, but, unfortunately, the technique may produce a query that, generally, is not equivalent to the original one. The conditions that a program must satisfy in order to preserve this equivalence are investigated.
- The proposed algorithm is an extension of (Greco, 2003) for Datalog$^\vee$ programs only. However, an in-depth analysis is also provided of the algorithm in (Greco, 2003), formally showing how it can work independently of the particular strategy adopted for simulating the binding occurring in top-down evaluation. In this way, this approach is orthogonal to the Magic-Set technique, since it can exploit any other rewriting strategy proposed in the literature.
- In order to verify the validity of this approach the technique has been tested with different disjunctive programs using the DLV system (Leone *et al.*, 2002). The experiments comparing the execution time required to answer the source and the optimized query, have achieved very encouraging results proving the value of the approach.

Even though there are a number of proposals for efficient query answering in disjunctive deductive databases under well-founded semantics, there are few effective techniques for top-down reasoning under stable model semantics. Thus, the paper also contributes to provide an effective, implemented technique for query answering under stable model semantics.

Finally, it should be stressed that even though the results in (Ullman, 1989a; Ullman, 1989b) suggest that this extension of the Magic-Set technique is equivalent to the form of binding propagation occurring in the top-down evaluation, the technique may suffer from some potential inefficiency due to the computation of additional predicates and rules needed in the rewriting (intrinsic in the Magic-Set technique). However, note that some redundancy can also occur in the top-down evaluation due to the tabling of intermediate results.

However, the main aim here is to formally prove the possibility of efficient query evaluation in bottom-up systems, breaking down the complexity of a brute force approach and still using the internal model generator. Thus, implementations of top-down systems supporting disjunction and stable model semantics have not been sought in order to make an experimental comparison w.r.t. this approach. This aspect calls for further study and research.

### 1.2 Related Work on the Evaluation of Disjunctive Programs

The (efficient) evaluation of disjunctive logic programs has a long history (Grant & Minker, 1986; Yahya & Henschen, 1985; Liu & Sunderraman, 1990; Yuan & Chiang, 1989). Recently there has been a growing interest in answering queries on disjunctive deductive databases. The two main approaches proposed in the literature for the evaluation of queries and programs are now discussed.

### 1.2.1 Bottom-Up Techniques

The definition of efficient bottom-up evaluation algorithms for assigning semantics to disjunctive deductive databases has been the subject of several proposals. In the following some of these approaches proposed in the literature are briefly described.

(Minker & Rajasekar, 1990) introduce the concept of *state*, consisting of a set of positive disjunctions, as the domain of a fixpoint operator that gives semantics to disjunctive logic programs. The fixpoint computation operates bottom-up and produces, as the resulting fixpoint, the *model state*, i.e. a state whose minimal models satisfy the disjunctive deductive database.

(Brass & Dix, 1994) propose a general approach for defining the semantics of disjunctive logic programs. The framework consists of: a semantical part, where the declarative meaning of a program is defined in an abstract way as the weakest semantics satisfying certain properties, and a procedural part, namely a bottom-up query evaluation method based on operators working on conditional facts. More specifically, the approach is based on the generation of a residual program, i.e. a program obtained by transforming the original one, which makes the use of disjunctive information explicit.

(Fernández & Minker, 1995a) introduce a new fixpoint characterization of the minimal models of disjunctive logic programs. The proposed operator, applied iteratively, is shown to characterize the perfect model semantics of stratified disjunctive logic programs. Based on these results the authors present a bottom-up evaluation algorithm for stratified disjunctive deductive databases that uses the model-tree data structure to both represent the information contained in the database and compute answers to queries.

(Leone *et al.*, 2002) propose the DLV system, which exploits an algorithm for computing stable models for disjunctive logic programs. This approach searches for stable models by using efficient fixpoint algorithms computing the semantics of programs. In particular, it obtains effective performances by using an (intelligent) ground instantiation of programs, i.e. a program in which unsatisfiable ground rules are deleted, and heuristics are used for implementing a backtracking search strategy for pruning the search space.

(Simons *et al.*, 2002) define a novel answer set programming language that generalizes normal logic programs. The language allows weighted constraint rules which increase the expressivity of the language to express different types of constraint (e.g. cardinality conditions) optimization capabilities. The declarative semantics extends the one for normal programs while the complexity of computing stable models for this novel language is comparable to that of normal programs (without considering optimizations).

### 1.2.2 Top-Down Techniques

Top-down query evaluation is based on refutation procedures. The first of such refutation techniques, called *SLD-resolution*, was introduced in (Kowalski, 1974) and is suitable for Datalog programs only, i.e. programs without negation and disjunction. (Clark, 1978) extended *SLD-resolution* to *SLDNF-resolution*, introducing the negation as a failure rule for inferring negative information.

In order to prevent the possibility of infinite loops and a large amount of redundant sub-derivations (intrinsic in *SLD-resolution*), the *SLG-resolution* was introduced (Chen & Warren, 1993; Chen & Warren, 1996). It is a tabling mechanism for the evaluation of the well founded semantics of logic programs (without disjunction), and it is the evaluation strategy underlying XSB, the best known state-of-the-art top-down tabling system (Sagonas *et al.*, 1994). (Shen *et al.*, 2002) present an optimization of *SLG-resolution*, called *SLT-resolution*.

The problem of top-down computation for disjunctive well-founded semantics (DWFS) was investigated in (Wang, 2001). Specifically, the author proposes a top-down procedure for disjunctive well-founded semantics called *D-SLS-resolution*, which can eventually be optimized by employing some techniques such as the tabling method of (Chen & Warren, 1996). A top-down method for testing DWFS membership, based on the characterization of the DWFS in terms of Gelfond-Lifschitz transformations, is presented in (Johnson, 2001).

A bottom-up procedure computing queries in a top-down fashion has also been proposed for minimal model semantics (Yahya, 2000; Yahya, 2002). The approach, suitable for positive queries, is based on the *duality* principle for interpreting logical connectives. In more detail, the duality transformation is obtained by reversing the direction of the implication arrows in the clauses representing both the program and the negation of the query goal. The application of a generic bottom-up procedure to the transformed clause set results in a top-down query answering.

Finally, (Johnson, 1999) shows that disjunctive stable models can be characterized in terms of cyclic covers, and in particular that such covers provide a powerful technique for characterizing the properties of query processing, compilation and view updating. A (correct and terminating) top-down method for query processing under the disjunctive stable model semantics is also presented. Supported covers have also been used to facilitate the top-down query processing under the possible model semantics (Johnson, 2002).

### *1.3 Plan of the Paper*

The paper is organized as follows: Section 2 presents preliminary definitions and results on disjunctive Datalog, minimal and stable model semantics and Magic-Set rewriting; Section 3 presents the Magic-Set rewriting for positive disjunctive queries; Section 4 extends the binding propagation technique to disjunctive queries with constraints; Section 5 presents experimental results showing the validity of the proposed technique; finally, Section 6 presents the conclusions.

## 2 Preliminaries

In this section standard concepts on Datalog, Magic-set rewriting, query equivalence, disjunction and constraints are reviewed.

### *2.1 Disjunctive Datalog*

The existence of alphabets of constants, variables and predicate symbols are aaumed. A *term* is a constant or a variable. An *atom* is of the form $p(t_1, \cdots, t_k)$ where $p$ is a $k$-ary predicate symbol and $t_1, \cdots, t_k$ are terms. A *literal* is an atom $A$ or its negation $\neg A$. A *Datalog$^{\vee,\neg}$* (or simply *disjunctive Datalog*) rule $r$ is a clause of the form:

$$a_1 \vee \cdots \vee a_m \leftarrow b_1, \cdots, b_k, \neg c_1, \cdots, \neg c_n$$

where $n, k, m \geq 0$, $n+k+m > 0$ and $a_1, \cdots, a_m, b_1, \cdots, b_k, c_1, \cdots, c_n$ are function-free atoms. The disjunction $a_1 \vee \cdots \vee a_m$ is called the *head* of $r$ and is denoted by $Head(r)$ while the conjunction $b_1, \cdots, b_k, \neg c_1, \cdots, \neg c_n$ is called the *body* and is denoted by $Body(r)$. If $m = 1$, then $r$ is *normal* (i.e. $\vee$-free) or *Datalog$^\neg$*; if $n = 0$, then $r$ is *positive* (i.e. $\neg$-free) or *Datalog$^\vee$*; if both $m = 1$ and $n = 0$, then $r$ is *normal and positive* or *Datalog*; if $k = n = 0$ $r$ is a *fact*, whereas if $m = 0$ $r$ is a *constraint* or *denial rule*, i.e. a rule which is satisfied only if $Body(r)$ is false. A *Datalog$^{\vee,\neg}$* program $\mathcal{P}$ is a finite set of Datalog$^{\vee,\neg}$ rules; it is *normal* (resp. *positive*) if all its rules are normal (resp. positive). Given a program $\mathcal{P}$ and a predicate symbol $g$ occurring in $\mathcal{P}$, the definition of $g$, denoted by $def(g, \mathcal{P})$, consists of all rules in $\mathcal{P}$ having $g$ in their heads.

The *Herbrand Universe* $U_\mathcal{P}$ of a program $\mathcal{P}$ is the set of all constants appearing in $\mathcal{P}$, and its *Herbrand Base* $B_\mathcal{P}$ is the set of all ground atoms constructed from the predicates appearing in $\mathcal{P}$ and the constants from $U_\mathcal{P}$. A *ground* term (resp. an atom, a literal, a rule or a program) is a term (resp. an atom, a literal, a rule or a program) where no variables occur in it. A rule $r'$ is a *ground instance* of a rule $r$, if $r'$ is obtained from $r$ by replacing every variable in $r$ with some constant in $U_\mathcal{P}$. We denote by $ground(\mathcal{P})$ the set of all ground instances of the rules in $\mathcal{P}$.

Given a set of ground atoms $\mathcal{I}$, a program $\mathcal{P}$, a predicate symbol $g$ and an atom $g(t)$, $\mathcal{I}[g]$ denotes the set of $g$-atoms in $\mathcal{I}$, whereas $\mathcal{I}[\mathcal{P}]$ denotes the set of atoms in $\mathcal{I}$ whose predicate symbol appears in the head of some rule of $\mathcal{P}$. Given a set of interpretations $S$, then $S[g] = \{M[g] | M \in S\}$ and $S[\mathcal{P}] = \{M[\mathcal{P}] | M \in S\}$.

An *interpretation* of $\mathcal{P}$ is any subset of $B_\mathcal{P}$. The value of a ground atom $L$ w.r.t. an interpretation $I$, $value_I(L)$, is 1 (*true*) if $L \in I$ and 0 (*false*) otherwise. The value of a ground negated literal *not* $L$ is $1 - value_I(L)$. The truth value of a conjunction of ground literals $C = L_1, \ldots, L_n$ is the minimum over the values of the $L_i$, i.e. $value_I(C) = min(\{value_I(L_i) \mid 1 \leq i \leq n\})$, while the value of a disjunction $D = L_1 \vee \cdots \vee L_n$ is its maximum, i.e. $value_I(D) = max(\{value_I(L_i) \mid 1 \leq i \leq n\})$; if $n = 0$, then $value_I(C) = true$ and $value_I(D) = false$. A ground rule $r$ is *satisfied* by $I$ if $value_I(Head(r)) \geq value_I(Body(r))$. Thus, a rule $r$ with empty body is satisfied by $I$ if $value_I(Head(r)) = true$. An interpretation $M$ for $\mathcal{P}$ is a model of $\mathcal{P}$ if $M$ satisfies all rules in $ground(\mathcal{P})$.

The *model-theoretic semantics* for a positive program $\mathcal{P}$ assigns the set of its *minimal models* $\mathcal{MM}(\mathcal{P})$. A model $M$ for $\mathcal{P}$ is minimal, if no proper subset of $M$ is a model for $\mathcal{P}$ (Minker, 1994). Accordingly, the program $\mathcal{P} = \{a \vee b \leftarrow\}$ has the two minimal models $\{a\}$ and $\{b\}$, i.e. $\mathcal{MM}(\mathcal{P}) = \{\ \{a\},\ \{b\}\ \}$. The more general *disjunctive stable model semantics* generalizes stable model semantics, previously defined for normal programs (Gelfond & Lifschitz, 1988) and also applies to programs with (unstratified) negation (Gelfond & Lifschitz, 1991; Przymusinski, 1991).

Let $\mathcal{P}$ be a logic program $\mathcal{P}$ and let $I$ be an interpretation for $\mathcal{P}$, $\mathcal{P}^I$ denotes the ground positive program derived from $ground(\mathcal{P})$ by (1) removing all rules that contain a negative literal $\neg a$ in the body and $a \in I$, and (2) removing all negative literals from the remaining rules. An interpretation $M$ is a (disjunctive) stable model for $\mathcal{P}$ if and only if $M \in \mathcal{MM}(\mathcal{P}^M)$.

For general $\mathcal{P}$, the stable model semantics assigns to $\mathcal{P}$ the set $\mathcal{SM}(\mathcal{P})$ of its *stable models*. It is well known that stable models are minimal models (i.e. $\mathcal{SM}(\mathcal{P}) \subseteq \mathcal{MM}(\mathcal{P})$) and that for negation-free programs minimal and stable model semantics coincide (i.e. $\mathcal{SM}(\mathcal{P}) = \mathcal{MM}(\mathcal{P})$) and that Datalog programs have a unique minimal model.

Predicate symbols can be either extensional (i.e. defined by the ground facts of a database — *EDB predicate symbols*), also called *base predicates*, or intensional (i.e. defined by the rules of the program — *IDB predicate symbols*), also called *derived predicates*. Thus a database $D$ consists of a set of ground facts having in the head a base predicate symbol (i.e. ground normal rules with empty body defining base predicates), whereas a program $\mathcal{P}$ consists of a set of (disjunctive) rules having in the heads derived predicate symbols.

A disjunctive Datalog *query* over a database defines a mapping from the database to a finite (possibly empty) set of finite (possibly empty) relations for the goal. A query is a pair $\langle G, \mathcal{P} \rangle$ where $G$ is an atom, called a *goal*, and $\mathcal{P}$ is a program. The application of a query $\mathcal{Q}$ to a database $D$ is denoted by $\mathcal{Q}(D)$ and the union of the program $\mathcal{P}$ and the facts in $D$ is denoted by $\mathcal{P}_D$. Clearly, all models for $\mathcal{P}_D$ contain the database $D$.

The result of a query $\mathcal{Q} = \langle G, \mathcal{P} \rangle$ on an input database $D$ is defined in terms of the stable models of $\mathcal{P}_D$, by taking either the union (*possible inference*) or the intersection (*certain inference*) of all models. Thus, given a program $\mathcal{P}$ and a database $D$, a ground atom $G$ is true, under possible (brave) semantics, if there exists a stable model $M$ for $\mathcal{P}_D$ such that $G \in M$. Analogously, $G$ is true, under certain (cautious) semantics, if $G$ is true in every stable model for $\mathcal{P}_D$.

Given an atom $G$ and an interpretation $M$, $A(G, M)$ denotes the set of substitutions for the variables in $G$ such that $G$ is true in $M$. The answer to a query $\mathcal{Q} = \langle G, \mathcal{P} \rangle$ over a database $D$ under *possible* (resp. *certain*) semantics, denoted $Ans_p(\mathcal{Q}, D)$ (resp., $Ans_c(\mathcal{Q}, D)$) is the relation $\cup_M A(G, M)$ (resp., $\cap_M A(G, M)$) such that $M \in \mathcal{SM}(\mathcal{P}, D)$. Two queries $\mathcal{Q}_1 = \langle G_1, \mathcal{P}_1 \rangle$ and $\mathcal{Q}_2 = \langle G_2, \mathcal{P}_2 \rangle$ are said to be *equivalent* under semantics $s$ ($\mathcal{Q}_1 \equiv_s \mathcal{Q}_2$) if for every database $D$ (on a fixed schema) $Ans_s(\mathcal{Q}_1, D) = Ans_s(\mathcal{Q}_2, D)$. Moreover, we say that two programs $\mathcal{P}_1$ and $\mathcal{P}_2$ are equivalent under a given semantics $s$: $\mathcal{P}_1 \equiv_s \mathcal{P}_2$ if for every atom $g$ $\langle g, \mathcal{P}_1 \rangle \equiv_s \langle g, \mathcal{P}_2 \rangle$. Finally, if $\mathcal{Q}_1 \equiv_p \mathcal{Q}_2$ and $\mathcal{Q}_1 \equiv_c \mathcal{Q}_2$ (the two queries or programs are equivalent under both brave and cautious semantics) we simply write $\mathcal{Q}_1 \equiv \mathcal{Q}_2$.

### 2.2 Magic-Set rewriting

In the literature different approaches have been proposed for the efficient bottom-up evaluation of queries, e.g. the Magic-Set (Bancilhon *et al.*, 1986), the supplementary Magic-Set (Beeri & Ramakrisnhan, 1991) and other specialized rewriting techniques (Greco *et al.*, 1995; Ullman, 1989a; Ramakrisnhan *et al.*, 1993). The key idea of all these techniques consists in the rewriting of deductive rules with respect to the query goal to answer the query without actually computing irrelevant facts. In this section we recall the *Magic-Set* rewriting is recalled, which is a general and well-known technique for the optimization of Datalog queries. Although the Magic-Set technique can be applied to general Datalog queries, for the sake of simplicity, here the technique for linear programs is presented, i.e. programs whose rules contain, at most, one body predicate mutually recursive with the head predicate.

The Magic-Set rewriting consists of three separate steps:

1. An *Adornment step* in which the relationship between a bound argument in the rule head and the bindings in the rule body is made explicit.
2. A *Generation step* in which the adorned program is used to generate the *magic rules* which simulate the top-down evaluation scheme.
3. A *Modification step* in which the adorned rules are modified by the magic rules generated in Step (2); these rules will be called *modified rules*.

An *adorned program*, $\mathcal{P}^\beta$ is a program whose predicate symbols have associated a string $\alpha$, defined on the alphabet $\{b, f\}$, of length equal to the arity of the predicate. A character $b$ (resp. $f$) in the i-th position of the adornment associated with a predicate $p$ means that the i-th argument of $p$ is bound (resp. free).

The adornment step consists in generating a new program whose predicates are adorned. Given a rule $r$ and an adornment $\alpha$ of the rule head, the adorned version of $r$ is derived as follows:

1. identify the distinctive arguments of the rules as follows: an argument is distinctive if it is bound in the adornment $\alpha$, is a constant or appears in a base predicate of the rule-body which includes an adornment argument;
2. assume that the distinctive arguments are bound and use this information in the adornment of the derived predicates in the rule body.

Adornments containing only $f$ symbols can be omitted.

Given a query $\mathcal{Q} = \langle q(T), \mathcal{P} \rangle$ and letting $\alpha$ be the adornment associated with $q(T)$, the set of adorned rules for $\mathcal{Q}$ is generated by 1) first computing the adorned version of the rules defining $q$ and 2) then generating, for each new adorned predicate $p^\alpha$ introduced in the previous step, the adorned version of the rules defining $p$ w.r.t. $\alpha$; Step 2 is repeated until no new adorned predicate is generated.

The second step of the process consists in using the adorned program for the generation of the magic rules. For each of the adorned predicates in the body of the adorned rule:

1. eliminate all the derived predicates in the rule body which are not mutually recursive with the rule head;
2. replace the derived predicate symbol $p^\alpha$ with *magic_$p^\alpha$* and eliminate the variables which are free w.r.t. $\alpha$;
3. Replace the head predicate symbol $q^\alpha$ with *magic_$q^\alpha$* and eliminate the variables which are free w.r.t. $\alpha$;
4. interchange the transformed head and derived predicate in the body.

Finally, the modification step of an adorned rule is performed as follows: for each adorned rule whose head is $p^\alpha(X)$, where $X$ is a list of variables, extend the rule body with *magic_$p^\alpha(X')$* where $X'$ is the list of variables in $X$ which are bound w.r.t. $\alpha$.

The final program will contain only the rules which are needed to answer the query.

*Example 1*

Consider the query $\mathcal{Q}_3 = \langle \mathtt{p(1,C)}, \mathcal{P}_3 \rangle$ where $\mathcal{P}_3$ is defined as follows:

$$\begin{array}{lcl}
\mathtt{p(X,C)} & \leftarrow & \mathtt{q(X,2,C)}. \\
\mathtt{q(X,Y,C)} & \leftarrow & \mathtt{a(X,Y,C)}. \\
\mathtt{q(X,Y,C)} & \leftarrow & \mathtt{b(X,Y,Z,W),\ q(Z,W,D),\ c(D,C)}.
\end{array}$$

The adorned version of $\mathcal{P}_3$ w.r.t. the query goal $\mathtt{p(1,C)}$ is:

$$\begin{array}{lcl}
\mathtt{p^{bf}(X,C)} & \leftarrow & \mathtt{q^{bbf}(X,2,C)}. \\
\mathtt{q^{bbf}(X,Y,C)} & \leftarrow & \mathtt{a(X,Y,C)}. \\
\mathtt{q^{bbf}(X,Y,C)} & \leftarrow & \mathtt{b(X,Y,Z,W),\ q^{bbf}(Z,W,D),\ c(D,C)}.
\end{array}$$

The rewritten query is $\mathcal{Q}'_3 = \langle \mathtt{p^{bf}(1,C)}, \mathcal{P}'_3 \rangle$ where $\mathcal{P}'_3$ is as follows:

$$\begin{array}{l}
\mathtt{magic\_p^{bf}(1)}. \\
\mathtt{magic\_q^{bbf}(X,2) \leftarrow magic\_p^{bf}(X)}. \\
\mathtt{magic\_q^{bbf}(Z,W) \leftarrow magic\_q^{bf}(X,Y), b(X,Y,Z,W)}. \\[4pt]
\mathtt{p^{bf}(X,C)} \leftarrow \mathtt{magic\_p^{bf}(X),\ q^{bbf}(X,2,C)}. \\
\mathtt{q^{bbf}(X,Y,C) \leftarrow magic\_q^{bbf}(X,Y),\ a(X,Y,C)}. \\
\mathtt{q^{bbf}(X,Y,C) \leftarrow magic\_q^{bbf}(X,Y), b(X,Y,Z,W), q^{bbf}(Z,W,D), c(D,C)}.
\end{array}$$

Note that the first set of rules consists of the magic rules generated in the second step, while the second set of rules consists of the modified rules. □

Observe that, although the technique presented here applies only to negation free linear programs, the Magic-Set rewriting is general and can also be applied to non-linear programs with some form of negation (e.g. stratified negation) where bindings are also propagated through derived predicates (Beeri & Ramakrisnhan, 1991).

Let $\mathcal{Q} = \langle G, \mathcal{P} \rangle$ be a query, then $Magic(\mathcal{Q})$ denotes the query derived from $\mathcal{Q}$ by applying the Magic-Set technique. The query $Magic(\mathcal{Q})$ will also be denoted as $\langle G^\alpha, magic(G, \mathcal{P}) \rangle$ where $G^\alpha$ denotes the adorned version of $G$, and $magic(G, \mathcal{P})$ denotes the rewriting of $\mathcal{P}$ w.r.t. the goal $G$. The rewritten program consists of two distinct sets of rules: a set of new rules (generated in Step (2)), called *magic rules*, and the set of *modified rules*, (generated in Step (3)) which is derived from the set of rules in the source program. The adorned rules generated in Step (1) are denoted by $Adorn(G, \mathcal{P})$.

## 3 Binding Propagation for Positive Queries

In this section the technique proposed in (Greco, 1999; Greco, 2003) for applying the Magic-Set technique to positive disjunctive Datalog programs is reviewed and extended. The technique proposed in (Greco, 1999; Greco, 2003) produces a rewriting of the source query which, under the bottom-up evaluation, simulates the propagation of bindings occurring in the query-goal, performed in the top-down evaluation.

It should be pointed out that this formalization stems from the approach in (Greco, 2003). However, it is further extended by providing a new result on query equivalence (not stated in (Greco, 2003)), which is crucial for allowing the technique to work in the case of disjunctive programs with constraint rules. This result is important and states that the rewriting technique is independent of the particular strategy adopted for simulating the propagation of bindings carried out in top-down evaluation. Therefore, even though conceptually introduced as an extension of the Magic-Set technique, this approach is orthogonal to the Magic-Set technique, since it can use any other rewriting strategy proposed in the literature.

For the sake of simplicity, the following running example is considered.

*Example 2*
Consider the query $\langle \texttt{ancestor(john, Y)}, \texttt{ANC} \rangle$ where the program `ANC` consists of the following rules:

```
father(X, Y) ∨ brother(X, Y)   ←   related(X, Y).
ancestor(X, Y)                 ←   father(X, Y).
ancestor(X, Y)                 ←   father(X, Z), ancestor(Z, Y).
```

The predicate `ancestor` defines the transitive closure of `father`, while `father` is defined by a disjunctive rule. □

Given a positive disjunctive Datalog program $\mathcal{P}$, the first step is to construct a suitable normal program.

*Definition 1*

Let $\mathcal{P}$ be a (positive) disjunctive Datalog program. The *extended standard* version of $\mathcal{P}$, denoted $esv(\mathcal{P})$, is the Datalog program derived from $\mathcal{P}$ by replacing each disjunctive rule $a_1 \vee \cdots \vee a_m \leftarrow B$ with

1. $m$ rules of the form $a_i \leftarrow B$   for $1 \leq i \leq m$, and
2. $m \times (m-1)$ rules of the form $a_i \leftarrow a_j, B$   for $1 \leq i, j \leq m$ and $i \neq j$.

Given a query $\mathcal{Q} = \langle G, \mathcal{P} \rangle$, we denote with $esv(\mathcal{Q})$ the query $\langle G, esv(\mathcal{P}) \rangle$.   □

*Example 3*

The program $esv(\texttt{ANC})$, where $\texttt{ANC}$ is the program presented in Example 2, consists of the following rules:

$$
\begin{array}{lcl}
\texttt{father}(X, Y) & \leftarrow & \texttt{related}(X, Y). \\
\texttt{brother}(X, Y) & \leftarrow & \texttt{related}(X, Y). \\
\texttt{father}(X, Y) & \leftarrow & \texttt{brother}(X, Y), \texttt{related}(X, Y). \\
\texttt{brother}(X, Y) & \leftarrow & \texttt{father}(X, Y), \texttt{related}(X, Y). \\
\texttt{ancestor}(X, Y) & \leftarrow & \texttt{father}(X, Y). \\
\texttt{ancestor}(X, Y) & \leftarrow & \texttt{father}(X, Z), \texttt{ancestor}(Z, Y).
\end{array}
$$

□

Observe that the rules introduced by applying Item 2 of Definition 1 are subsumed by those introduced by applying Item 1; indeed, the semantics of $esv(\mathcal{P})$ is not affected by the insertion of these rules. However, these rules are necessary in order to allow the propagation of the bindings, as will be clear in the following.

It should be pointed out that programs $\mathcal{P}$ and $esv(\mathcal{P})$ are not equivalent; in fact, $\mathcal{P}$ is a disjunctive Datalog program that is able to express all the queries in $\Sigma_2^P$, while $esv(\mathcal{P})$ is a positive Datalog program that is able to express a subset of the queries computable in polynomial time.

The second step is to derive a program that must be equivalent to the original one. Let us first present some notation.

*Definition 2*

Given a (positive) disjunctive Datalog program $\mathcal{P}$, $ESV(\mathcal{P})$ denotes the program derived from $esv(\mathcal{P})$ by replacing each derived predicate symbol $g$ with a new predicate symbol $G$. Given a query $\mathcal{Q} = \langle g(t), \mathcal{P} \rangle$, $ESV(\mathcal{Q})$ denotes the query $\langle G(t), ESV(\mathcal{P}) \rangle$ where $G$ is the new symbol used to replace $g$.   □

*Definition 3*

Let $\mathcal{P}$ be a (positive) disjunctive Datalog program. The *restricted* version of $\mathcal{P}$, denoted by $RV(\mathcal{P})$, is the disjunctive Datalog program defined as follows:

$$RV(\mathcal{P}) = \{a_1 \vee \cdots \vee a_m \leftarrow A_1, \cdots, A_m, B \mid a_1 \vee \cdots \vee a_m \leftarrow B \in \mathcal{P}\}$$

where each $A_i$, for $1 \leq i \leq m$, is the atom replacing $a_i$ in the program $ESV(\mathcal{P})$. The rewritten version of $\mathcal{P}$ is $Rew(\mathcal{P}) = RV(\mathcal{P}) \cup ESV(\mathcal{P})$. Given a query $\mathcal{Q} = \langle g(t), \mathcal{P} \rangle$, $Rew(\mathcal{Q})$ denotes the query $\langle g(t), Rew(\mathcal{P}) \rangle$. □

Observe that in the above definition, the rewriting of the program $\mathcal{P}$ does not take into account the goal. This is not true for rewriting techniques propagating bindings.

*Example 4*

The program `Rew(ANC)`, where `ANC` is the program defined in Example 2, consists of the following rules $RV(\mathtt{ANC})$:

$$
\begin{array}{rcl}
\mathtt{father(X,Y) \vee brother(X,Y)} & \leftarrow & \mathtt{FATHER(X,Y), BROTHER(X,Y), related(X,Y)}. \\
\mathtt{ancestor(X,Y)} & \leftarrow & \mathtt{ANCESTOR(X,Y), father(X,Y)}. \\
\mathtt{ancestor(X,Y)} & \leftarrow & \mathtt{ANCESTOR(X,Y), father(X,Z), ancestor(Z,Y)}.
\end{array}
$$

plus the set of rules $ESV(\mathtt{ANC})$ :

$$
\begin{array}{rcl}
\mathtt{FATHER(X,Y)} & \leftarrow & \mathtt{related(X,Y)}. \\
\mathtt{BROTHER(X,Y)} & \leftarrow & \mathtt{related(X,Y)}. \\
\mathtt{FATHER(X,Y)} & \leftarrow & \mathtt{BROTHER(X,Y), related(X,Y)}. \\
\mathtt{BROTHER(X,Y)} & \leftarrow & \mathtt{FATHER(X,Y), related(X,Y)}. \\
\mathtt{ANCESTOR(X,Y)} & \leftarrow & \mathtt{FATHER(X,Y)}. \\
\mathtt{ANCESTOR(X,Y)} & \leftarrow & \mathtt{FATHER(X,Z), ANCESTOR(Z,Y)}.
\end{array}
$$

The rewritten query is $\langle \mathtt{ancestor(john,Y)}, \mathtt{Rew(ANC)} \rangle$    □

The programs $\mathcal{P}$ and $Rew(\mathcal{P})$ actually have the same semantics (w.r.t. the predicates in $\mathcal{P}$).

*Proposition 1*

Let $\mathcal{P}$ be a (positive) disjunctive Datalog program, and $Rew(\mathcal{P})$ be the rewritten version of $\mathcal{P}$. Then, for every atom $g(t)$, $\langle g(t), \mathcal{P} \rangle \equiv \langle g(t), Rew(\mathcal{P}) \rangle$.

**Proof.** In order to prove that for any $g(t)$, $\langle g(t), \mathcal{P} \rangle \equiv \langle g(t), Rew(\mathcal{P}) \rangle$, under both possible and certain semantics, it will be shown that for any database $D$, an interpretation $M$ of $Rew(\mathcal{P})_D$ is a stable model if, and only if, $M[\mathcal{P}_D]$ is a stable model for $\mathcal{P}_D$. Moreover, only minimal models can be considered, since for disjunctive positive Datalog programs, the set of stable models of a given program coincides with the set of minimal models.

It should be pointed out that the program $Rew(\mathcal{P})$ consists of two distinct components: i) the program $ESV(\mathcal{P})$ whose rules only depend on $D$, and ii) the program $RV(\mathcal{P})$ whose rules depend on predicates defined in $ESV(\mathcal{P})$ and $D$. Hence, the set of stable models of $Rew(\mathcal{P})_D$ can be computed in a level wise manner.

As for the models of $ESV(\mathcal{P})_D$, observe that $ESV(\mathcal{P})_D$ has a unique minimal model, since it is a positive program. Let $M_{ESV}$ be such a model, and let $M_{esv} = \{a \mid A \in M_{ESV}\}$.

Then, $\mathcal{MM}(ESV(\mathcal{P})_D \cup RV(\mathcal{P})_D) = \mathcal{MM}(FM_{ESV} \cup RV(\mathcal{P})_D)$, where $FM_{ESV}$

denotes the set of facts associated with each atom in the model $M_{ESV}$. Finally, in order to conclude the proof, it can be claimed that $M_{ESV} \cup M$ is a minimal model for $\mathcal{MM}(FM_{ESV} \cup RV(\mathcal{P})_D)$ if, and only if, $M$ is a minimal model for $\mathcal{P}_D$.

In fact, consider a rule $r : a_1 \vee \cdots \vee a_m \leftarrow b, b_1, \cdots, b_n$ in $\mathcal{P}$ and the corresponding rule $r' : a_1 \vee \cdots \vee a_m \leftarrow A_1, \cdots, A_m, b, b_1, \cdots, b_n$ in $RV(\mathcal{P})$, where $b$ is a conjunction of derived predicates, while $b_1, \cdots, b_n$ are extensional predicates. It is easy to see, that for any model $M$ for $\mathcal{P}_D$, we have $M \subseteq M_{esv}$. Hence, if $b, b_1, \cdots, b_n$ is true in $M$ then also $B, b_1, \cdots, b_n$ is true in $M_{ESV}$. Then, due to the rules $A_i \leftarrow B, b_1, \cdots, b_n$ in $ESV(\mathcal{P})$, $A_1, \cdots, A_m$ is true in $M_{ESV}$, too.

Conversely, if $b$ is false in $M$, then the body of $r'$ is trivially false (no matter what the evaluation of $B$), too. Hence, after the assertion of the facts in $M_{ESV}$, the semantics of $RV(\mathcal{P})_D$ is not affected from $r'$ the predicates $A_1, \cdots, A_m$ are removed. □

The previous proposition states that the program $\mathcal{P}$ and $Rew(\mathcal{P})$, obtained by restricting the rules of $\mathcal{P}$, are equivalent. In fact, adding the atoms defined in $ESV(\mathcal{P})$ inside the body of the rules of $\mathcal{P}$, does not make any effective restriction, as for every ground atom $a(t)$ appearing in the head of a disjunctive ground rule $r$ there is a ground atom $A(t)$ which is derived from $ESV(\mathcal{P})$.

Thus, in the following, instead of using the program $ESV(\mathcal{P})$ to restrict the rules in $\mathcal{P}$, a different program is considered which makes an effective and sound restriction; this program will be obtained by performing the binding propagation from the query goal into the rules of the program $esv(\mathcal{P})$. However, program $ESV(\mathcal{P})$ is used instead of program $esv(\mathcal{P})$, in order to distinguish between atoms of the source program and atoms making restrictions.

Throughout the paper, the Magic-Set technique is considered which is a well-known and general technique. For special classes of queries we could use specialized techniques as the choice of the rewriting technique is independent and orthogonal w.r.t. the proposed framework.

For details about the Magic-Set technique the reader should refer to (Beeri & Ramakrisnhan, 1991; Ullman, 1989a) while for specialized rewriting techniques see (Ullman, 1989a; Ramakrisnhan *et al.*, 1993; Greco *et al.*, 1995).

Let's now formally provide a way of "collecting" all the adorned predicates generated by the rewriting of queries.

*Definition 4*
Given a (positive) disjunctive Datalog program $\mathcal{P}$, a predicate symbol $p$ appearing in $\mathcal{P}$ and an adorned program $\mathcal{P}^\beta$ derived from $\mathcal{P}$, then

$$Coll(p, \mathcal{P}, \mathcal{P}^\beta) = \{p(X_1, \cdots, X_k) \leftarrow p^\alpha(X_1, \cdots, X_k) | \text{for every } p^\alpha \text{ in } \mathcal{P}^\beta \text{ derived from } p\}$$

denotes the set of rules (also called *collecting rules*) used to collect the atoms of the predicates having different adornments, but the same predicate symbols.

Moreover,

$$Coll(\mathcal{P}, \mathcal{P}^\beta) = \bigcup_{p \text{ appearing in } \mathcal{P}} Coll(p, \mathcal{P}, \mathcal{P}^\beta)$$

denotes the set of all collecting rules derived from $\mathcal{P}$ and $\mathcal{P}^\beta$. □

With respect to the rewriting of Definition 3, $ESV(\mathcal{P})$ is now replaced by the rules of the Magic-Set program $\mathcal{P}' = Magic(G(t), ESV(\mathcal{P}))$ (where $G$ is the symbol corresponding to $g$ in $ESV(\mathcal{P})$) plus the rules used to collect the atoms with different adornments of $\mathcal{P}'$, denoted by $Coll(ESV(\mathcal{P}), \mathcal{P}')$.

*Example 5*

Consider the program ANC of Example 4 and the query goal ancestor(john, Y). The corresponding extended standard query is $ESV(\langle \text{ancestor}(\text{john}, Y), \text{ANC} \rangle) = \langle \text{ANCESTOR}(\text{john}, Y), ESV(\text{ANC}) \rangle$. The first step consists in the generation of adornments for derived predicates. From the propagation of bindings only one binding for the predicate $\text{ANCESTOR}^b(\text{john}, Y)$ is derived.

Thus, the program $Magic(\text{ANCESTOR}(\text{john}, Y), ESV(\text{ANC}))$ is as follows:

```
FATHER^bf(X,Y)        ← magic_FATHER^bf(X), related(X,Y).
BROTHER^bf(X,Y)       ← magic_BROTHER^bf(X), related(X,Y).
FATHER^bf(X,Y)        ← magic_FATHER^bf(X), BROTHER^bf(X,Y), related(X,Y).
BROTHER^bf(X,Y)       ← magic_BROTHER^bf(X), FATHER^bf(X,Y), related(X,Y).
ANCESTOR^bf(X,Y)      ← magic_ANCESTOR^bf(X), FATHER^bf(X,Y).
ANCESTOR^bf(X,Y)      ← magic_ANCESTOR^bf(X), FATHER^bf(X,Z), ANCESTOR^bf(Z,Y).

magic_ANCESTOR^bf(john).
magic_ANCESTOR^bf(Z)  ← magic_ANCESTOR^bf(X), FATHER^bf(X,Z).
magic_FATHER^bf(X)    ← magic_ANCESTOR^bf(X).
magic_BROTHER^bf(X)   ← magic_FATHER^bf(X).
magic_FATHER^bf(X)    ← magic_BROTHER^bf(X).
```

Here the predicate $\text{magic\_ANCESTOR}^{bf}$ computes all ancestors which are relevant to establish whether a given person is an ancestor of john.

The set $Coll(ESV(\text{ANC}), Magic(\text{ANCESTOR}(\text{john}, Y), ESV(\text{ANC})))$, consisting of the rules collecting atoms with the same predicate and different adornments, is

```
ANCESTOR(X,Y)  ← ANCESTOR^bf(X,Y).
FATHER(X,Y)    ← FATHER^bf(X,Y).
BROTHER(X,Y)   ← BROTHER^bf(X,Y).
```

These rules collect into ANCESTOR (resp. FATHER, BROTHER) all the ANCESTOR (resp. FATHER, BROTHER) atoms with different adornments. Since there is only one adornment for each predicate, adornments and collecting rules could be eliminated. □

*Definition 5*

Let $\mathcal{Q} = \langle g(t), \mathcal{P} \rangle$ be a disjunctive Datalog query, then the disjunctive Magic-Set rewriting of $\mathcal{P}$ w.r.t. $g(t)$, denoted by $Disj\_Magic(g(t), \mathcal{P})$, is the following program: $RV(\mathcal{P}) \cup Coll(ESV(\mathcal{P}), Magic(G(t), ESV(\mathcal{P}))) \cup Magic(G(t), ESV(\mathcal{P}))$. □

*Example 6*

The complete rewriting of the program in Example 2 consists of the rules in $Coll(ESV(\texttt{ANC}), Magic(\texttt{ANCESTOR}(\texttt{john}, \texttt{Y}), ESV(\texttt{ANC}))) \cup Magic(\texttt{ANCESTOR}(\texttt{john}, \texttt{Y}), ESV(\texttt{ANC}))$ (shown in Example 5) plus the rules in $RV(\texttt{ANC})$:

$$\texttt{father}(\texttt{X}, \texttt{Y}) \vee \texttt{brother}(\texttt{X}, \texttt{Y}) \leftarrow \texttt{FATHER}(\texttt{X}, \texttt{Y}), \texttt{BROTHER}(\texttt{X}, \texttt{Y}), \texttt{related}(\texttt{X}, \texttt{Y}).$$
$$\texttt{ancestor}(\texttt{X}, \texttt{Y}) \leftarrow \texttt{ANCESTOR}(\texttt{X}, \texttt{Y}), \texttt{father}(\texttt{X}, \texttt{Y}).$$
$$\texttt{ancestor}(\texttt{X}, \texttt{Y}) \leftarrow \texttt{ANCESTOR}(\texttt{X}, \texttt{Y}), \texttt{father}(\texttt{X}, \texttt{Z}), \texttt{ancestor}(\texttt{Z}, \texttt{Y}). \quad \square$$

From the above observation and definition, combined with Proposition 1 the following result proved in (Greco, 2003) can be derived.

*Fact 1*

Let $\mathcal{Q} = \langle g(t), \mathcal{P} \rangle$ be a Datalog$^\vee$ query, then, $\mathcal{Q} \equiv \langle g(t), Disj\_Magic(g(t), \mathcal{P}) \rangle$. $\quad \square$

## 4 Binding Propagation in Datalog$^\vee$ Programs with Constraints

This section formally introduces a technique for propagating bindings into Datalog$^\vee$ queries with *strong* or *classical* constraints, a simple and powerful form of unstratified negation. A strong constraint is a rule with empty head of the form: $\leftarrow B(X)$ where $B(X)$ is a conjunction of literals and $X$ is a vector of range restricted variables, which must be satisfied in each model[1].

Contrary to standard Datalog, where bindings are propagated from the head of rules into the body, the problem with programs containing constraints is that the bindings need to be propagated also through the constraints. For instance, if one is interested in knowing whether $p(1)$ is true in a program where there is the constraint $\leftarrow p(X), q(X)$, then the truth value of $q(1)$ also needs to be evaluated.

Thus, the truth value of each ground atom in a constraint depends on the truth value of the other atoms appearing in the same ground constraint, and, hence, in a more abstract perspective, constraints behave in a similar manner as disjunctive rules when propagating bindings into their heads.

In the following, a Datalog$^\vee$ program $\mathcal{P}$ with constraints (Datalog$^{\vee, \leftarrow}$) will be denoted by a pair $(\mathcal{P}_R, \mathcal{P}_C)$, where $\mathcal{P}_R$ is a nonempty set of (positive) disjunctive rules and $\mathcal{P}_C$ is a set of constraints. It is worth noting that $\mathcal{P}_R$ being a nonempty set of positive disjunctive rules, the only form of negation contained in $\mathcal{P}$ is that related to the rewriting of constraints. Moreover, the use of constraints instead of general (possibly unstratified) negation is not a limitation since Datalog$^{\vee, \leftarrow}$ has the same expressive power as Datalog$^{\vee, \neg}$. Indeed, since in (Eiter *et al.*, 1997a) it has been shown that Datalog$^{\vee, \neg_s}$ has the same expressivity as Datalog$^{\vee, \neg}$, whereas in (Zumpano, 2004) it has been shown that stratified negation can be simulated by using disjunction and constraints, we have that $\text{Datalog}^{\vee,\leftarrow} \equiv \text{Datalog}^{\vee,\neg_s}$.

---

[1] It should be recalled that, under stable models semantics, a constraint $\leftarrow B(X)$ could be rewritten into an equivalent rule with unstratified negation of the form: $p_i \leftarrow B(X), \neg p_i$, where $p_i$ is a new predicate symbol not defined elsewhere in the program.

As has been shown in (Zumpano, 2004), that every rule containing stratified negation can be rewritten in $\text{Datalog}^{\vee,\leftarrow}$ rules. For instance, consider the following stratified rule where predicate $b$ does not depend on predicate $p$:

$$p(X) \leftarrow a(X), \neg b(X)$$

This rule can be rewritten as:

$$\begin{aligned} 1: & \ p(X) \leftarrow p'(X) \\ 2: & \ p'(X) \vee b'(X) \leftarrow a(X) \\ 3: & \ \leftarrow p'(X), b'(X) \\ 4: & \ \leftarrow b'(X) \neg b(X) \\ 5: & \ \leftarrow p'(X), b(X) \end{aligned}$$

where the rule 2 together with the constraint 3 states that $a$ is partitioned into $p'$ and $b'$, whereas constraints 4 and 5 state, respectively, that $b'$ must be a subset of $b$ and that the intersection between $p'$ and $b$ must be empty. Therefore, tuples being in $a - b$ must be in $p'$. The rule 1 is necessary only in the case the predicate $p$ in the source program is defined by more than one rule. In the following we assume that a predicate symbol $p$ cannot appear both positively and negatively in two different constraints. This restriction does not make any limitation on the expressive power of the language as stratified negation can be emulated by considering the above restricted form of constraints.

By following the same guidelines as in the previous section, the technique and the main results, by using a running example.

*Example 7*

Suppose to have the query $\langle \texttt{2col(1,2)}, \texttt{Coloring} \rangle$ checking whether a graph is 3-colorable and whether the colors *red* and *blue* can be assigned to nodes 1 and 2 respectively. The program $\texttt{Coloring}$ consists of the following rules:

$$\begin{aligned} \texttt{r}: & \ \texttt{2col(X,Y)} \leftarrow \texttt{color(X,red), color(Y,blue)}. \\ \texttt{s}: & \ \texttt{color(X,red)} \vee \texttt{color(X,blue)} \vee \texttt{color(X,yellow)} \leftarrow \texttt{node(X)}. \\ \texttt{c}: & \ \leftarrow \texttt{edge(X,Y), color(X,C), color(Y,C)}. \end{aligned}$$

□

*Definition 6*

Given a set of constraints $P_C$, $esv(P_C)$ denotes the set of Datalog rules obtained by replacing each constraint in $P_C$ having the form

$$\leftarrow a_1, \cdots, a_k, b_1, \cdots, b_m, \neg c_1, \cdots, \neg c_n$$

where $a_1, \cdots, a_k$ are base atoms, $b_1, \cdots, b_m$ are derived atoms and $c_1, \cdots, c_n$ are negated literals (either base or derived), with the following set of rules:

$$b_i \leftarrow b_1, \cdots, b_{i-1}, b_{i+1}, \cdots, b_m, a_1, \cdots, a_k, \neg c_1, \cdots, \neg c_n \quad \forall i \in [1 \cdots m]$$

Given program $\mathcal{P} = (\mathcal{P}_R, \mathcal{P}_C)$, the *extended standard version* of $\mathcal{P}$, denoted by $esv(\mathcal{P}) = esv((\mathcal{P}_R, \mathcal{P}_C)) = esv(\mathcal{P}_R) \cup esv(\mathcal{P}_C)$.

□

*Example 8*

The extended standard version of the program `Coloring` of Example 7, obtained by the rewriting of constraints and disjunctive rules and denoted by $esv(\texttt{Coloring})$, is as follows:

$\texttt{r}_1: \texttt{2col(X,Y)} \leftarrow \texttt{color(X,red), color(Y,blue)}.$

$\texttt{s}_1: \texttt{color(X,red)} \leftarrow \texttt{node(X)}.$
$\texttt{s}_2: \texttt{color(X,blue)} \leftarrow \texttt{node(X)}.$
$\texttt{s}_3: \texttt{color(X,yellow)} \leftarrow \texttt{node(X)}.$

$\texttt{s}_4: \texttt{color(X,red)} \leftarrow \texttt{color(X,blue), node(X)}.$
$\texttt{s}_5: \texttt{color(X,blue)} \leftarrow \texttt{color(X,red), node(X)}.$
$\texttt{s}_6: \texttt{color(X,red)} \leftarrow \texttt{color(X,yellow), node(X)}.$
$\texttt{s}_7: \texttt{color(X,yellow)} \leftarrow \texttt{color(X,red), node(X)}.$
$\texttt{s}_8: \texttt{color(X,blue)} \leftarrow \texttt{color(X,yellow), node(X)}.$
$\texttt{s}_9: \texttt{color(X,yellow)} \leftarrow \texttt{color(X,blue), node(X)}.$

$\texttt{c}_1: \texttt{color(X,C)} \leftarrow \texttt{edge(X,Y), color(Y,C)}.$
$\texttt{c}_2: \texttt{color(Y,C)} \leftarrow \texttt{edge(X,Y), color(X,C)}.$

where the rule $\texttt{r}_1$ is derived from $\texttt{r}$, rules $\texttt{s}_1 - \texttt{s}_9$ are derived from the rule $\texttt{s}$ and rules $\texttt{c}_1$ and $\texttt{c}_2$ are derived from constraint $\texttt{c}$. □

As in the case of disjunctive programs without constraints, $ESV(\mathcal{P})$ denotes the program derived from $esv(\mathcal{P})$ by replacing each derived predicate symbol $g$ with a new predicate symbol $G$.

*Definition 7*

Let $\mathcal{P} = (\mathcal{P}_R, \mathcal{P}_C)$ be a Datalog$^\vee$ program with constraints. Program $Rew(\mathcal{P})$ is defined as $RV(\mathcal{P}_R) \cup \mathcal{P}_C \cup ESV(\mathcal{P})$. Given a query $\mathcal{Q} = \langle g(t), \mathcal{P} \rangle$, $Rew(\mathcal{Q})$ denotes the query $\langle g(t), Rew(\mathcal{P}) \rangle$. □

Notice that, with respect to programs without constraints, $RV(\mathcal{P}_R)$ is replaced by $RV(\mathcal{P}_R) \cup \mathcal{P}_C$ and $ESV(\mathcal{P}_R)$ by $ESV(\mathcal{P}_R) \cup ESV(\mathcal{P}_C)$.

*Example 9*

The set of restricted rules in $RV(\texttt{Coloring})$, derived from the program `Coloring` of Example 7, is as follows:

$\texttt{2col(X,Y)} \leftarrow \texttt{2COL(X,Y), color(X,red), color(Y,blue)}.$
$\texttt{color(X,red)} \vee \texttt{color(X,blue)} \vee \texttt{color(X,yellow)} \leftarrow \texttt{COLOR(X,red), COLOR(X,blue),}$
$\texttt{COLOR(X,yellow), node(X)}.$

where predicates `2COL` and `COLOR` are defined in $ESV(\texttt{Coloring})$ which is derived from program $esv(\texttt{Coloring})$, shown in Example 8, by replacing `2col` with `2COL` and `color` with `COLOR`. The complete rewriting of the program `Coloring` consists of the above rules plus the constraint:

$\leftarrow \texttt{edge(X,Y), color(X,C), color(Y,C)}.$

and the rules in $ESV(\texttt{Coloring})$, presented in Example 8. □

The first interesting result is that (as for Datalog$^\vee$ programs without constraints) the above rewriting method does not affect the semantics of the query.

*Proposition 2*

Let $\mathcal{P} = (\mathcal{P}_R, \mathcal{P}_C)$ be a Datalog$^\vee$ program with constraints, and $Rew(\mathcal{P})$ be the rewritten version of $\mathcal{P}$. Then, for every atom $g(t)$, $\langle g(t), \mathcal{P} \rangle \equiv \langle g(t), Rew(\mathcal{P}) \rangle$.

**Proof.** Recall that $Rew(\mathcal{P}) = Rew((\mathcal{P}_R, \mathcal{P}_C)) = RV(\mathcal{P}_R) \cup \mathcal{P}_C \cup ESV(\mathcal{P})$; hence $Rew(\mathcal{P}) = Rew(\mathcal{P}_R) \cup \mathcal{P}_C \cup ESV(\mathcal{P}_C)$ can also be written.

Let us now consider program $\mathcal{P}' = Rew(\mathcal{P}_R) \cup ESV(\mathcal{P}_C)$; using the same arguments as for Proposition 1, $\langle g(t), \mathcal{P}_R \rangle \equiv \langle g(t), \mathcal{P}' \rangle$ is derived. In fact, it has already been shown that adding the atoms defined in $ESV(\mathcal{P}_R)$ inside the body of the rules of $\mathcal{P}$, does not make any effective restriction, provided that $\mathcal{P}_R$ is a positive program; moreover, program $ESV(\mathcal{P}_C)$ is also a positive program that possibly enlarges the unique model $M_{ESV}$ of $ESV(\mathcal{P})$, without affecting the models of $RV(\mathcal{P}_R)$. Finally, the result follows by observing that the set of constraints $\mathcal{P}_C$ affects the result of the query only in the case when there is some ground constraint that is not satisfied, and by the fact that $\mathcal{P}$ and $Rew(\mathcal{P})$ share the same set of constraints. □

Thus, a viable way for reducing the number of models to be computed is to consider a suitable rewriting of $ESV(\mathcal{P})$, which is able to make an effective and sound restriction by simulating the binding propagation occurring in top-down evaluation. This rewriting is carried out by means of the Magic-Set technique, which in fact limits attention to the models that are really needed for answering the query.

*Definition 8*

Let $\mathcal{Q} = \langle g(t), \mathcal{P} \rangle$ with $\mathcal{P} = (\mathcal{P}_R, \mathcal{P}_C)$, then the disjunctive Magic-Set rewriting of $\mathcal{P}$ w.r.t. $g(t)$, denoted by $Disj\_Magic(g(t), \mathcal{P})$, is defined as follows:
$RV(\mathcal{P}_R) \cup \mathcal{P}_C \cup Coll(ESV(\mathcal{P}), Magic(G(t), ESV(\mathcal{P}))) \cup Magic(G(t), ESV(\mathcal{P}))$. □

*Example 10*

Consider again the query $\langle \texttt{2col}(1,2), \texttt{Coloring} \rangle$ of Example 7.

The program $Magic(\texttt{2COL}(1,2), ESV(\texttt{Coloring}))$, obtained by applying the Magic-Set technique to the query $ESV(\langle \texttt{2col}(1,2), \texttt{Coloring} \rangle)$, is as follows:

```
magic_2COLbb(1, 2).
magic_COLORbb(X, red)     ←  magic_2COLbb(X, Y).
magic_COLORbb(Y, blue)    ←  magic_2COLbb(X, Y).
magic_COLORbb(X, blue)    ←  magic_COLORbb(X, red).
magic_COLORbb(X, red)     ←  magic_COLORbb(X, blue).
magic_COLORbb(X, yellow)  ←  magic_COLORbb(X, red).
magic_COLORbb(X, red)     ←  magic_COLORbb(X, yellow).
magic_COLORbb(X, blue)    ←  magic_COLORbb(X, yellow).
magic_COLORbb(X, yellow)  ←  magic_COLORbb(X, blue).
magic_COLORbb(Y, C)       ←  magic_COLORbb(X, C), edge(X, Y).
magic_COLORbb(X, C)       ←  magic_COLORbb(Y, C), edge(X, Y).
```

$$\begin{aligned}
\text{COLOR}^{\text{bb}}(\text{X},\text{red}) &\leftarrow \text{magic\_COLOR}^{\text{bb}}(\text{X},\text{red}), \text{node}(\text{X}).\\
\text{COLOR}^{\text{bb}}(\text{X},\text{blue}) &\leftarrow \text{magic\_COLOR}^{\text{bb}}(\text{X},\text{blue}), \text{node}(\text{X}).\\
\text{COLOR}^{\text{bb}}(\text{X},\text{yellow}) &\leftarrow \text{magic\_COLOR}^{\text{bb}}(\text{X},\text{yellow}), \text{node}(\text{X}).\\
\\
\text{COLOR}^{\text{bb}}(\text{X},\text{red}) &\leftarrow \text{magic\_COLOR}^{\text{bb}}(\text{X},\text{red}), \text{COLOR}^{\text{bb}}(\text{X},\text{blue}).\\
\text{COLOR}^{\text{bb}}(\text{X},\text{blue}) &\leftarrow \text{magic\_COLOR}^{\text{bb}}(\text{X},\text{blue}), \text{COLOR}^{\text{bb}}(\text{X},\text{red}).\\
\text{COLOR}^{\text{bb}}(\text{X},\text{red}) &\leftarrow \text{magic\_COLOR}^{\text{bb}}(\text{X},\text{red}), \text{COLOR}^{\text{bb}}(\text{X},\text{yellow}).\\
\text{COLOR}^{\text{bb}}(\text{X},\text{yellow}) &\leftarrow \text{magic\_COLOR}^{\text{bb}}(\text{X},\text{yellow}), \text{COLOR}^{\text{bb}}(\text{X},\text{red}).\\
\text{COLOR}^{\text{bb}}(\text{X},\text{blue}) &\leftarrow \text{magic\_COLOR}^{\text{bb}}(\text{X},\text{blue}), \text{COLOR}^{\text{bb}}(\text{X},\text{yellow}).\\
\text{COLOR}^{\text{bb}}(\text{X},\text{yellow}) &\leftarrow \text{magic\_COLOR}^{\text{bb}}(\text{X},\text{yellow}), \text{COLOR}^{\text{bb}}(\text{X},\text{blue}).\\
\\
\text{COLOR}^{\text{bb}}(\text{X},\text{C}) &\leftarrow \text{magic\_COLOR}^{\text{bb}}(\text{X},\text{C}), \text{edge}(\text{X},\text{Y}), \text{COLOR}^{\text{bb}}(\text{Y},\text{C}).\\
\text{COLOR}^{\text{bb}}(\text{Y},\text{C}) &\leftarrow \text{magic\_COLOR}^{\text{bb}}(\text{Y},\text{C}), \text{edge}(\text{X},\text{Y}), \text{COLOR}^{\text{bb}}(\text{X},\text{C}).\\
\\
\text{2COL}^{\text{bb}}(\text{X},\text{Y}) &\leftarrow \text{magic\_2COL}^{\text{bb}}(\text{X},\text{Y}), \text{COLOR}^{\text{bb}}(\text{X},\text{red}), \text{COLOR}^{\text{bb}}(\text{Y},\text{blue}).
\end{aligned}$$

while the rules in $Coll(ESV(\texttt{Coloring}), Magic(\text{2COL}(1,2), ESV(\texttt{Coloring})))$ are:

$$\begin{aligned}
\text{2COL}(\text{X},\text{Y}) &\leftarrow \text{2COL}^{\text{bb}}(\text{X},\text{Y}).\\
\text{COLOR}(\text{X},\text{red}) &\leftarrow \text{COLOR}^{\text{bb}}(\text{X},\text{red}).\\
\text{COLOR}(\text{X},\text{blue}) &\leftarrow \text{COLOR}^{\text{bb}}(\text{X},\text{blue}).\\
\text{COLOR}(\text{X},\text{yellow}) &\leftarrow \text{COLOR}^{\text{bb}}(\text{X},\text{yellow}). \quad \square
\end{aligned}$$

Before formally presenting the correctness of the rewriting technique, let us resume the process and make some comments. The rewriting process takes in input a query $Q = \langle g(t), \mathcal{P} \rangle = \langle g(t), (\mathcal{P}_R, \mathcal{P}_C) \rangle$ and first generates the equivalent query $Q' = \langle g(t), (RV(\mathcal{P}_R) \cup ESV(\mathcal{P}), \mathcal{P}_C) \rangle$, where $ESV(\mathcal{P})$ is a normal program. Next the query $Q'' = \langle g(t), (RV(\mathcal{P}_R) \cup \mathcal{P}', \mathcal{P}_C) \rangle$, where $\mathcal{P}'$ is the optimized program derived from $ESV(\mathcal{P})$ is produced. To answer the source query $Q$ is sufficient to consider the minimal models of $\mathcal{P}_R$ which satisfies $\mathcal{P}_C$. As the program $ESV(\mathcal{P})$ (resp. $\mathcal{P}'$) may contain negated literals, to answer the rewritten query $Q'$ (resp. $Q''$) the stable models of $RV(\mathcal{P}_R) \cup ESV(\mathcal{P})$ (resp. $RV(\mathcal{P}_R) \cup \mathcal{P}'$) satisfying $\mathcal{P}_C$ have to be computed. Moreover, under the assumption that a predicate symbol $p$ cannot appear both positively and negatively in two different constraints, the program $ESV(\mathcal{P})$ is stratified and therefore has a unique stable model (namely the perfect or stratified model). Therefore, in order to answer the query $Q'$ it is sufficient to compute the perfect model $M$ of $ESV(\mathcal{P})$ and then to compute the minimal models of $RV(\mathcal{P}_R) \cup M$ satisfying $\mathcal{P}_C$. For unstratified $ESV(\mathcal{P})$ the complete set of stable models has to be considered. However, as already mentioned, problems in the second level of the polynomial hierarchy can be expressed by means of Datalog$^\vee$ programs with restricted constraints.

### 4.1 Query Equivalence Results

For the sake of presentation in Figure 1 the main steps provided by the whole algorithm presented in this section are explicitly pointed out. The algorithm takes as input a query $\langle g(t), \mathcal{P} \rangle$ and a database $D$, and outputs the set of stable models of $Disj\_Magic(g(t), \mathcal{P})_D$; obviously, this set can be used for answering the query under both the possible and certain semantics.

```
Input: A query Q = ⟨g(t), P⟩ with P = (P_R, P_C), a database D;
Output: The stable models of (Disj_Magic(g(t), P))_D.
var: ESV, RV, Coll: set of rules;
begin
   let ESV := ∅, RV := ∅, Coll := ∅;
 ⎧  //*** generation of the restricted version ***
 ⎨  for each rule r ∈ P_R of the form a_1(t) ∨ ... ∨ a_n(t) ← B do
 ⎩     insert a_1 ∨ ··· ∨ a_n ← A_1, ···, A_n, B into RV;
 ⎧  //*** generation of the extended standard version ***
 ⎪  for each rule r ∈ P_R of the form a_1(t) ∨ ... ∨ a_n(t) ← B do begin
 ⎪     insert a_i ← B    for 1 ≤ i ≤ n into ESV;
 ⎨     insert a_i ← a_j, B    for 1 ≤ i, j ≤ n and i ≠ j into ESV;
 ⎪  end
 ⎪  for each constraint c ∈ P_C of the form ← a_1, ···, a_k, b_1, ··, b_m, ¬c_1, ···, ¬c_n do
 ⎩     insert b_i ← b_1, ··, b_{i−1}, b_{i+1}, ··, b_k, a_1, ···, a_k, ¬c_1, ···, ¬c_n for 1 ≤ i ≤ m into ESV;
 ⎧  //*** application of the Magic-Set technique to a normal program ***
 ⎨  let Magic := Magic(G(t), ESV(P));
 ⎧  //*** generation of the collecting rules ***
 ⎪  for each predicate symbol p defined in P with arity k do
 ⎨     for each adornment α of p in Magic do
 ⎩        insert p(X_1, ··, X_k) ← p^α(X_1, ··, X_k) into Coll;
   return SM(RV ∪ P_C ∪ Coll ∪ Magic ∪ D);
end.
```

Fig. 1.  **Algorithm** Extended Magic-Set (*Magic_Partial*)

For Datalog$^\vee$ programs without constraints the correctness of Algorithm in Fig. 1 follows from Fact 1. For Datalog$^\vee$ program with constraints, its correctness (under proper assumptions) will be provided below.

First of all observe that as in the case of disjunctive queries without constraints, the application of the Magic-Set technique to queries with constraints produces a query that can be evaluated more efficiently. Unfortunately, for Datalog$^{\vee,\leftarrow}$ queries the technique previously described produces a query that, generally, is not equivalent to the original one. This result is due to the fact that Magic-Set technique not only focuses on the models really relevant for answering the query, but also computes part of the models and not models in their entirety. In contrast, constraints express conditions that must hold for every ground instance of the program including atoms which are not relevant to the query. This observation will be clearer after the following example.

*Example 11*
Consider the query ⟨`2col(1,2)`, `Coloring`⟩ applied to the program of Example 8 on the graph shown in Figure 2 consisting of two disconnected components, say $C_1$ and $C_2$. Since both nodes 1 and 2 belong to component $C_1$, there is no way for propagating bindings from 1 and 2 into the component $C_2$. This means that the query goal only depends on component $C_1$ and the colorability of $C_2$ affects the result only in the case when the component $C_2$ is not colorable.    □

The above example suggests that the original query $\mathcal{Q} = \langle g(t), \mathcal{P} \rangle$ and the query

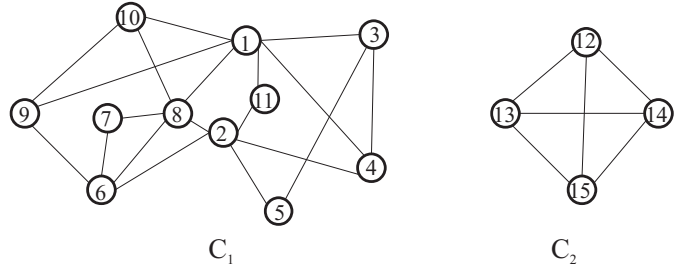

Fig. 2. A graph with two components, with $C_2$ being not 3-Colorable.

$\mathcal{Q}'$, obtained by applying the Magic-Set technique to $\mathcal{Q}$, are equivalent if the program $\mathcal{P}_D$ admits stable models. Before formally proving such an intuition, some preliminary definitions are provided.

Given a Datalog$^\vee$ program with constraints $\mathcal{P}$ and an interpretation $N$ for $\mathcal{P}$, then $\mathcal{P}/N$ denotes the set of rules in $ground(\mathcal{P})$ which are true w.r.t. $N$, and $\overline{\mathcal{P}/N}$ denotes the set of rules in $ground(\mathcal{P})$ which are false w.r.t. $N$, i.e. $\overline{\mathcal{P}/N} = ground(\mathcal{P}) - \mathcal{P}/N$.

In order to capture the meaning of the application of the Magic-Set technique in the case of disjunctive rules with constraints, use is made of the following concept.

### Definition 9
Let $\mathcal{P}$ be a Datalog$^\vee$ program with constraints and $D$ a database and let $M$ be an interpretation for $\mathcal{P}_D$, then $M$ is a *pre-model* for $\mathcal{P}_D$ if there exists $N \in \mathcal{SM}(\mathcal{P}_D)$ such that $M \subseteq N$.   □

In the case of Datalog$^\vee$ programs without constraints, every model of the program rewritten by means of the Magic-Set technique can be extended to be a model of the original program, i.e. given a program $\mathcal{P}_D$, every model of $Disj\_Magic(g(t), \mathcal{P}_D)$ restricted to the predicates in $\mathcal{P}_D$ is a pre-model for $\mathcal{P}_D$.

### Lemma 2
Let $\langle g(t), \mathcal{P}_R \rangle$ be a disjunctive Datalog query without constraints. Then, for any database $D$, $M \in \mathcal{SM}(Disj\_Magic(g(t), (\mathcal{P}_R)_D))$ if, and only if, there exists $N \in \mathcal{SM}(P_D)$ such that $M[(\mathcal{P}_R)_D] \subseteq N$.

**Proof.** From Fact 1 it is known that $\langle g(t), \mathcal{P}_R \rangle \equiv \langle g(t), Disj\_Magic(g(t), \mathcal{P}_R) \rangle$. Indeed, for any model $M \in \mathcal{SM}(Disj\_Magic(g(t), (\mathcal{P}_R)_D))$ the program $\overline{(\mathcal{P}_R)_D/M[(\mathcal{P}_R)_D]}$ consists of a set of ground rules that are false in $(\mathcal{P}_R)_D$ only because they are not necessary for answering the query, i.e. they are not used for propagating the binding. Thus, there exists a way for extending model $M[(\mathcal{P}_R)_D]$ into a new model $N$ for $(\mathcal{P}_R)_D$ that satisfies the above rules.   □

The above lemma states that from the rewritten program atoms not inferable from the source program cannot be inferred, apart from those introduced for binding propagation and to collect atoms with different adornments.

In the case of a program $\mathcal{P}$ with constraint rules, the above observation does not hold as it is not always the case that for a given program $\mathcal{P}_D$ every model of $Disj\_Magic(g(t), \mathcal{P}_D)$ restricted to the predicates in $\mathcal{P}_D$ is a pre-model for $\mathcal{P}_D$.

*Lemma 3*

*There exists a query $\langle g(t), \mathcal{P} \rangle$, with $\mathcal{P} = (\mathcal{P}_R, \mathcal{P}_C)$ and $\mathcal{P}_C \neq \emptyset$, a database $D$ and a model $M$ in $\mathcal{SM}(Disj\_Magic(g(t), \mathcal{P}_D))$ such that $M[\mathcal{P}_D]$ is not a pre-model for $\mathcal{P}_D$.*

**Proof.** Consider the query $\langle \texttt{2col(1,2)}, \texttt{Coloring} \rangle$ presented in Example 7. Consider the database $D$ modelling the graph in Example 11.

Then, $\mathcal{P}_D$ does not have stable models, since the graph is not 3-colorable. However, $\mathcal{SM}(Disj\_Magic(2col(1,2), \mathcal{P}_D)) \neq \emptyset$, since the component of the graph containing both nodes 1 and 2 is 3-colorable — any such legal coloring corresponds in fact to a stable model. □

Hence, the Magic-Set technique (and any other similar technique to propagate bindings) for Datalog$^\vee$ programs with constraints does not produce a query equivalent to the original one. Nonetheless, it is natural to investigate some restrictions that may guarantee the soundness and/or the completeness of the answers.

*Theorem 1*

*Let $\mathcal{Q} = \langle g(t), \mathcal{P} \rangle$ be a query, where $\mathcal{P} = (\mathcal{P}_R, \mathcal{P}_C)$ is Datalog$^\vee$ program with constraints, and $D$ a database. Then, for each model $M'$ of $\mathcal{P}_D$, there exists $M \in \mathcal{SM}(Disj\_Magic(g(t), \mathcal{P}_D))$, with $M[\mathcal{P}_D] \subseteq M'$ being a pre-model for $\mathcal{P}_D$.*

**Proof.** Recall that $Disj\_Magic(g(t), (\mathcal{P}_R, \mathcal{P}_C))$ is defined as $RV(\mathcal{P}_R) \cup \mathcal{P}_C \cup Coll(ESV(\mathcal{P}), Magic(G(t), ESV(\mathcal{P}))) \cup Magic(G(t), ESV(\mathcal{P}))$. Given two programs $S_1$ and $S_2$ such that $S_2 = RV(S_1)$, then we also denote $S_1$ as $RV^{-1}(S_2)$.

Let $\mathcal{P}' = Disj\_Magic(g(t), \mathcal{P})) - \mathcal{P}_C$. Then, $\langle g(t), \mathcal{P}' \rangle \equiv \langle g(t), \mathcal{P}_R \rangle \equiv \langle g(t), Disj\_Magic(g(t), \mathcal{P}_R) \rangle$ as the additional rules in $\mathcal{P}'$ do not make any restrictions on the ground rules in $\mathcal{P}_R$ used to derive the goal $g(t)$. The program $\mathcal{P}_1 = ground(\mathcal{P}')$ consists of the three distinct sets: $\mathcal{P}_{11} = ground(RV(\mathcal{P}_R))$ containing restricted rules, $\mathcal{P}_{12} = ground(Coll(ESV(\mathcal{P}), Magic(G(t), ESV(\mathcal{P}))))$ containing collecting rules and $\mathcal{P}_{13} = ground(Magic(G(t), ESV(\mathcal{P})))$ containing adorned rules.

Consider now program $\mathcal{P}_2 = ground(\mathcal{P}_R) - RV^{-1}(\mathcal{P}_{11})$ containing all the rules in $ground(\mathcal{P}_R)$ not relevant for the query goal $g(t)$. It is obvious that, letting $\mathcal{P}_3 = ground(\mathcal{P}_R) - \mathcal{P}_2$, we have that $\langle g(t), \mathcal{P}_3 \rangle \equiv \langle g(t), \mathcal{P}_1 \rangle$ as both contain all the ground rules relevant for the query goal.

Therefore, for any database $D$, $\mathcal{SM}(\mathcal{P}_D) = \mathcal{SM}(ground(\mathcal{P})_D) = \mathcal{SM}((\mathcal{P}_3 \cup \mathcal{P}_C)_D) \times \mathcal{SM}((\mathcal{P}_2 \cup \mathcal{P}_C)_D)$, where for any two sets of stable models $A$ and $B$, $A \times B = \{M_1 \cup M_2 \mid M_1 \in A \wedge M_2 \in B\}$.

Moreover, since $\langle g(t), \mathcal{P}_3 \cup \mathcal{P}_C \rangle \equiv \langle g(t), \mathcal{P}_1 \cup \mathcal{P}_C \rangle$, as the constraints act on atoms derived from both $\mathcal{P}_3$ and $\mathcal{P}_1$, and $\langle g(t), \mathcal{P}_1 \cup \mathcal{P}_C \rangle \equiv \langle g(t), \mathcal{P}' \cup \mathcal{P}_C \rangle$, we conclude that $\langle g(t), \mathcal{P}_3 \cup \mathcal{P}_C \rangle = \langle g(t), Disj\_Magic(g(t), \mathcal{P}_D) \rangle$.

Hence, given a model $M'$ of $\mathcal{P}_D$, there exist (i) $M \in \mathcal{SM}(Disj\_Magic(g(t), \mathcal{P}_D))$ and (ii) $M'' \in \mathcal{SM}((\mathcal{P}_2 \cup \mathcal{P}_C)_D)$ such that $M' = M[\mathcal{P}_D] \times M''$.  □

The above theorem can be restated in the following more explicative form.

*Corollary 1*
Let $\mathcal{Q}' = \langle g(t), \mathcal{P} = (\mathcal{P}_R, \mathcal{P}_C) \rangle$ be a query, $D$ be a database, and $\mathcal{Q} = \langle g(t), Disj\_Magic(g(t), \mathcal{P}) \rangle$ then

- $Ans_b(\mathcal{Q}', D) \subseteq Ans_b(\mathcal{Q}, D)$, and
- $Ans_c(\mathcal{Q}', D) \supseteq Ans_c(\mathcal{Q}, D)$

**Proof.** The relation $Ans_b(\mathcal{Q}', D) \subseteq Ans_b(\mathcal{Q}, D)$ straightforwardly derives from Theorem 1. For the answer under cautious semantics, we distinguish two cases. If $\mathcal{SM}(\mathcal{P}_D) = \emptyset$, then any ground atom is trivially in $Ans_c(\mathcal{Q}', D)$, and hence $Ans_c(\mathcal{Q}', D) \supseteq Ans_c(\mathcal{Q}, D)$. Assume, then, that $\mathcal{SM}(\mathcal{P}_D) \neq \emptyset$. In this case, $Ans_c(\mathcal{Q}', D) = \cap_{M'} A(g, M')$, where $A(g, M')$ denotes the set of substitutions for the variables in $g$ such that $g$ is true in $M'$, for each $M'$ stable model of $\mathcal{P}_D$. Since, for each stable model $M'$, there exists $M \in \mathcal{SM}(Disj\_Magic(g(t), \mathcal{P}_D)$, with $M[\mathcal{P}_D] \subseteq M'$, it follows that $Ans_c(\mathcal{Q}', D) = \cap_{M'} A(g, M') = \cap_M A(g, M[\mathcal{P}_D])$. As $Disj\_Magic(g(t), \mathcal{P}_D)$ might contain additional stable models w.r.t. those of $\mathcal{P}_D$, $Ans_c(\mathcal{Q}, D)$ is a subset of $\cap_M A(g, M[\mathcal{P}_D])$. Hence, we have that $Ans_c(\mathcal{Q}', D) = \cap_M A(g, M[\mathcal{P}_D]) \supseteq Ans_c(\mathcal{Q}, D)$.  □

The above result has shed some light into the effectiveness of the Magic-Set technique for disjunctive program with constraints. Indeed, the rewriting is both *bravely complete*, i.e., under the brave semantics it guarantees to compute all the answers for the original query, and *cautiously sound*, i.e., under the cautious semantics it guarantees that no false answers are in fact computed. Actually, since we cannot prove that soundness and completeness hold at the same time in any semantics, the algorithm presented in Figure 1 will be also called *Magic_Partial* algorithm.

A natural extension, called *Magic_Total* algorithm, is shown in Figure 3, and consists of a first application of the *Magic_Partial*, and in a successive evaluation of the stable models of the program in which the binding has not been propagated. It is worth noting that the *Magic_Total* algorithm returns the set of all the models; however, it is almost trivial to modify the algorithm, in order to implement the possible and certain semantics in a more direct and efficient way.

We conclude by observing that applying the algorithm *Magic_Total* (rather than *Magic_Partial*) produces an overhead in query answering. Then, in the next section the results of some experiments are presented which quantify this overhead.

## 5 Experimental results

In this section some experimental results are presented to give an idea of the improvements which can be obtained by means of this technique; the proposed algorithm has been included in the system presented in (Greco, 2003), and all the

```
Input: A disjunctive Datalog query Q = ⟨g(t), P⟩, a database D;
Output: The set of stable models of P_D in which g(t) is evaluated true;
var: M, MP: set of models;
begin
  let M := ∅;
  let MP := Magic_Partial (⟨g(t), P⟩, D);
  for each M ∈ MP do
    let M := M ∪ M[P_D] × SM(P_D/M[P_D]);
  return M;
end.
```

Fig. 3. **Algorithm** *Magic_Total*

experiments have been carried out by means of the DLV system (Leone *et al.*, 2002) on a PC with a Pentium 4, 1.7 GHz, 512 Mbyte of RAM under the operating system Linux.

It should be pointed out that this proposal is neither a new evaluation strategy nor a new implementation, but an optimization useful for efficiently evaluating bound queries in bottom-up engines; in fact, this contribution lies in having formally proved that the Magic-Set rewriting can also be extended to deal with constraint rules, and, hence, implementation issues are subject for further research.

Nonetheless, experimental evaluation makes sense, since some classical decision and optimization problems applied in "extreme" situations have been considered, representative of a wide spectrum of real cases, in which the improvements of our techniques are negligible and highly evident, respectively.

Note that the timings considered in all the following experimental results do not include the time for the rewriting. In any case, this time does not affect the overall result as an exponential speedup in the execution times is obtained between the source and optimized (rewritten) versions.

*SIMPLE EXAMPLES.* Consider disjunctive program $\mathcal{P}_1$ consisting of the following rule:

$$p(X) \vee q(X) \leftarrow a(X, Y)$$

and program $\mathcal{P}_2$ obtained by adding to $\mathcal{P}_1$ the constraint

$$\leftarrow p(X), a(X, Y), q(Y), X \leq 1.$$

Figure 4.(i) shows the results obtained by considering query $\langle \mathtt{p(1)}, \mathcal{P}_1 \rangle$, evaluated over the database $D$ consisting of a set of facts $a(1, 2), a(2, 3), \cdots, a(k, k+1)$. The figure presents the execution time for the source program and the optimized version, obtained by applying the rewriting technique for positive queries. The experiments have been performed with databases, whose number of facts is shown on the $x$-axis, while the $y$-axis shows the time taken to evaluate the query (in seconds). The improvement of the optimized query is extremely high (observe that the scale of the y-axis is logarithmic), and is due to the fact that the optimized version propagates

the binding of the query $p(1)$, with the effect of reducing the models to be computed from an exponential to a constant number.

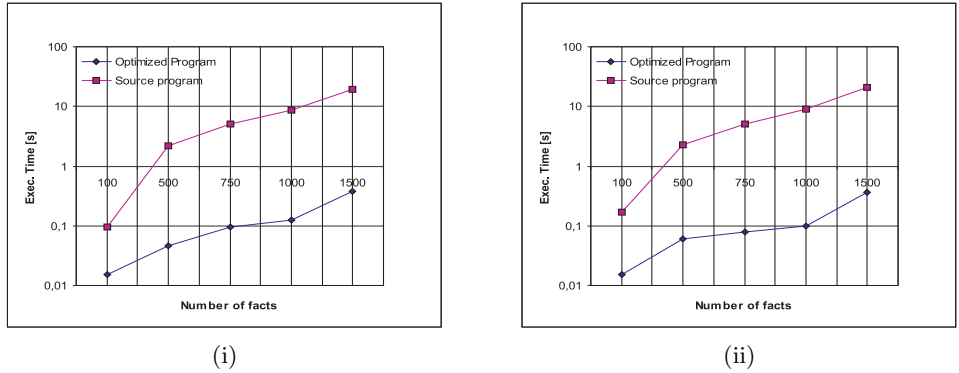

Fig. 4. Results for the query of (i) Example 1, and (ii) Example 2.

In Figure 4.(ii) the results obtained by evaluating query $\langle \mathtt{p(1)}, \mathcal{P}_2 \rangle$ are presented. By comparing the results shown in Figure 4.(i) and Figure 4.(ii) it can be observed that also in this case there is an exponential speedup between the optimized query and the source query, as the number of ground rules in the optimized version is still constant.

### 5.1 Search problems

For the following queries graphs have been used having the structure depicted in Figure 5 with *base = height* and output grade equal to 3 and 2, respectively. Here *base* denotes the number of nodes in the same layer, *height* the number of nodes in the same column and *grade* the number of arcs starting from each node not belonging to the top layer (or equivalently, the number of arcs ending at every node not belonging to the bottom layer). The number of nodes in the graph is $base \times height$, and the number of arcs is $(base-1) \times (height-1) \times grade + (base-1) + (height-1)$.

*3-COLORING.* The query of Example 7 is considered with input graphs consisting of two disconnected components with variable sizes; the results are shown in Fig. 6 where the computation of the source query and the computation of the query rewritten using both the *Magic_Partial* and *Magic_Total* algorithm have been considered. In particular, in 6.(i) the two components are of very different size, and the nodes in the query goal belong to the larger one. The graph shows the execution times as the size of the larger component changes. In 6.(ii) the same experiments repeated using two components with the same number of nodes are presented.

Note that in the first experiment, whose results are shown in Fig. 6.(i), there is no difference between the response time of the two optimized versions because the Magic-Set technique propagates the binding in the greatest component that dominates the size of the graph. For both optimized programs the advantage with respect to the source program is evident, since the size of the component on which the binding does not propagate is quite irrelevant. In Fig. 6.(ii) graphs with two components of the same size are being considered and, therefore, obtaining a full solution (i.e. Magic-Total) requires almost twice the time required for the partial solution (i.e. Magic-Partial).

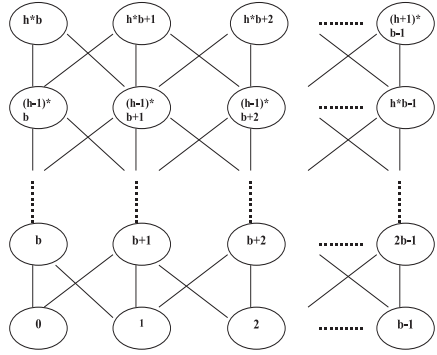

Fig. 5. Graph structures.

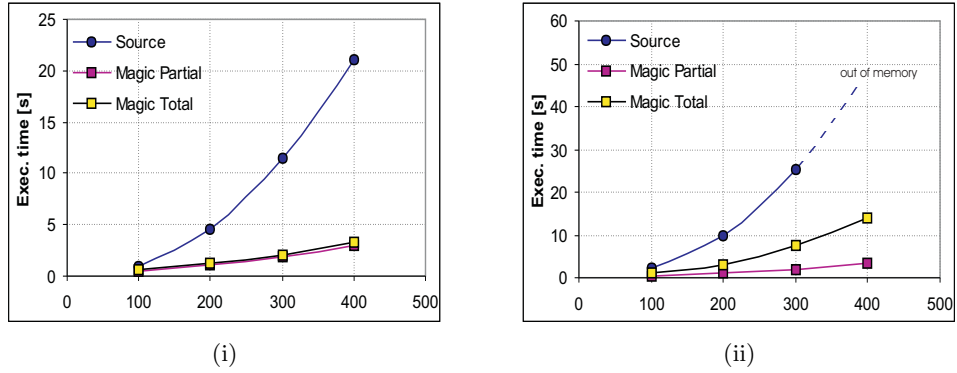

Fig. 6. Execution time for the 3-coloring problem.

*k-DOMINATING SET*. Given a graph $G = \langle V, E \rangle$, a subset of the vertex set $V' \subseteq V$ is a dominating set if for all $u \in V - V'$ there is a $v \in V'$ for which $(u, v) \in E$. The k-dominating set problem consists in finding a partition of the nodes into $V_1, \cdots, V_k$ disjoint dominating sets for $G$. The 3-Dominating Set problem, denoted by 3PDS, can be formalized by means of the following set of logic rules:

$$
\begin{aligned}
\texttt{v1(X)} \vee \texttt{nv1(X)} &\leftarrow \texttt{node(X)}. \\
\texttt{v2(X)} \vee \texttt{nv2(X)} &\leftarrow \texttt{node(X)}. \\
\texttt{v3(X)} \vee \texttt{nv3(X)} &\leftarrow \texttt{node(X)}. \\
&\leftarrow \texttt{v1(X)}, \texttt{v2(X)}. \\
&\leftarrow \texttt{v1(X)}, \texttt{v3(X)}. \\
&\leftarrow \texttt{v2(X)}, \texttt{v3(X)}.
\end{aligned}
$$

```
        ←   nv1(X), not connected1(X).
        ←   nv2(X), not connected2(X).
        ←   nv3(X), not connected3(X).

connected1(Y)   ←   v1(X), edge(X, Y).
connected2(Y)   ←   v2(X), edge(X, Y).
connected3(Y)   ←   v3(X), edge(X, Y).
```

Note that the first group of rules and the strong constraints induce a partition of the nodes into 3 disjoint sets, $v1$, $v2$ and $v3$.

The query $\langle (\texttt{v1}(1), \texttt{v2}(2), \texttt{v3}(3)), \texttt{3DS} \rangle$ is supplied on a graph $G$ consisting of two components $C_1$ and $C_2$. The size of $C_1$ is fixed and it is assumed that this component contains nodes 1, 2 and 3. In some experiments the size of $C_2$ was varied and the response time calculated for the source program described above, and the optimized program produced by the Algorithm in Fig. 1.

The results, presented in Fig. 7, show that the optimized program is not affected by the size of $C_2$ as there is no way of propagating the binding from nodes 1,2 and 3.

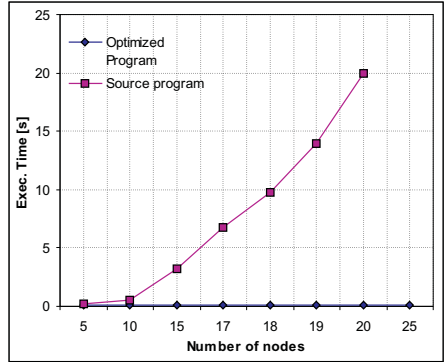

Fig. 7. Execution time for the 3-Dominating Set problem.

### 5.2 Optimization problems

In this section the possibility of using Magic-Set techniques for the optimization of queries over disjunctive Datalog programs is explored. In order to also express optimization problems the approach used in the DLV system is considered as well as the consideration, in addition to strong constraints, weak constraints. Weak constraints represent constraints which should be respected, but if they cannot be eventually enforced, then they only invalidate the portion of the program which they are concerned with (Greco, 1998). Therefore, strong constraints express a set of conditions that have to be satisfied, while weak constraints express a set of desiderable conditions that may be violated and their informal semantics is to minimize the number of violated instances.

A *weak constraint* is a rule of the form $\Leftarrow b_1, \cdots, b_k, \neg b_{k+1}, \cdots, \neg b_{k+m}$. Given a program $P \cup W$ where $P$ is a set of rules and $W$ a set of weak constraints, an interpretation $M$ is a stable model for $P \cup W$ if $M$ is a stable model for $P$ which minimizes the number of rules not satisfied in $ground(W)$. Thus, the stable models of $P$ can be ordered w.r.t. the number of weak constraints not satisfied in $ground(W)$; the preferred stable models are those which minimize this number.

Like strong constraints, weak constraints are not used to infer atoms, but only to check that the computed set verifies a given property. In (Buccafurri *et al.*, 2000; Greco, 1998) it is proved that the introduction of weak constraints allows the solution of optimization problems since each weak constraint can be regarded as an "objective function" of an optimization problem.

*Example 12*

Given a graph $G = \langle V, E \rangle$, defined by means of the unary predicate *node* and the binary predicate *edge*, we can model the MAX_CLIQUE problem, asking for the clique of $G$ having maximum size, by means of the following disjunctive Datalog program with both *strong* and *weak* constraints:

$$c(X) \vee nc(X) \leftarrow node(X).$$
$$\leftarrow c(X), c(Y), X \neq Y, not\ edge(X, Y).$$
$$\Leftarrow nc(X).$$

Note that the first rule is used for creating all possible partitions of nodes into c and nc, and the second one (i.e. the strong constraint) is used for ensuring that $c$ is a clique, i.e. each couple of nodes in the clique must be connected by an edge, while the weak constraint minimizes the number of vertices that are not in the clique, or equivalently it maximizes the size of $c$.

Consider the query $\langle c(1), \text{MAX\_CLIQUE} \rangle$ over the graph of Figure 2, asking whether node 1 belongs to a clique of maximum size. It is easy to see that nodes 1, 9 and 10 form a clique of size 3, which is also the size of the maximum clique in the component $C_1$. Component $C_2$ is a clique of size 4, and, hence, the above query is *false*. In contrast, the query $\langle c(12), \text{MAX\_CLIQUE} \rangle$ asking whether node 12 belongs to a clique of maximum size is evaluated *true*.

Now, observe that both the above queries are evaluated true, by using the Magic-Set rewriting; in fact, when we ask whether 1 belongs to a clique of maximum size, there is no way for propagating the binding in component $C_2$, where the maximum clique actually is. □

From the above example, it is clear that any query optimization technique applied in the presence of weak constraints, will eventually lead to a different semantics consisting in a 'local' optimization, rather than in a global one.

In many circumstances, this semantics can also be desirable. In all other circumstances, the global solution can be simply obtained by comparing the different solutions obtained while making local optimizations, by exploiting the same approach used in developing the *Magic_Total* algorithm.

The extension of the technique to deal also with weak constraints is outside the scope of this paper. Therefore, in the next two paragraphs some hints for future research are presented. In particular, two optimization problems are considered where the propagation of bindings defines a partition of the input graph $G$ into two separated components $G_1$ and $G_2$ and the optimal solution can be obtained by first computing an optimal solution using component $G_1$ and next finding the global optimal solution starting from the partial solution obtained in the first step and considering component $G_2$.

However, in many cases an optimization problem $Opt$ over a given graph $G$ can be defined by decomposing the graph $G$ into separated subgraphs $G_1, \cdots, G_k$ and to compute the $k$ subproblems. Thus, let $Opt(G_i, O_i)$ be the optimization problem over the (sub-)graph $G_i$ using the partial solution $O_i$, we say that $Opt$ is decomposable if $Opt(G, \emptyset) = Opt(G_k, O_k)$ where $O_1 = Opt(G_1, \emptyset)$ and $O_i = Opt(G_i, O_{i-1})$ for $i \in [2 \cdot \cdot k]$.

*MAX CLIQUE.* Let us consider the program of Example 12, and let us supply the query $\langle$`c(1)`, `MAX_CLIQUE`$\rangle$, by using a graph consisting of two components with the structure shown in Figure 5. The experimental results are shown in Fig. 8. In particular Fig. 8.(i) shows the execution time for a graph consisting of two components having the same size. In Fig. 8.(ii) the difference in the execution time between the source program and the optimized one for different sizes of the second component it is shown. Note that in Fig. 8.(ii), if the second component is empty (0 nodes) the source program performs a little better than the optimized one as the second program presents an overhead due to the instantiation and the computation of the magic rules. The advantage of using the optimized program becomes more evident as the size of the second component increases.

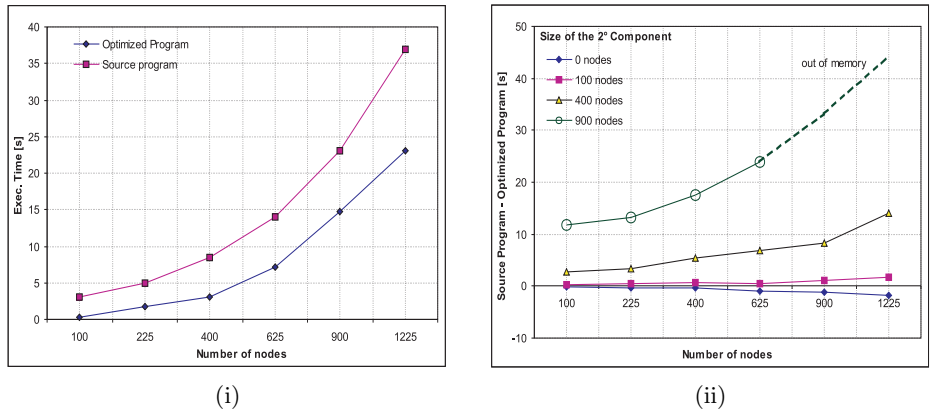

Fig. 8. The Max Clique Problem.

It is worth noting that the previous example is a prototype of the *guess* and *check*

paradigm, that, as already pointed out, has been proved to be the most intuitive way for expressing *NP* (optimization) problems.

*MIN COLORING.* Given a graph $G = \langle V, E \rangle$ a coloring for $G$ is minimum if, in the assignment of colors to vertices, it uses the minimum number of colors. The disjunctive Datalog program modelling the MIN_COLORING problem is the following:

$$\mathtt{col(X,I) \vee ncol(X,I)} \leftarrow \mathtt{node(X), color(I)}.$$

$$\leftarrow \mathtt{col(X,I), col(Y,I), edgeT(X,Y)}.$$
$$\leftarrow \mathtt{col(X,I), col(X,J), I\,!=\,J}.$$
$$\leftarrow \mathtt{node(X), not\ colored(X)}.$$

$$\mathtt{colored(X)} \leftarrow \mathtt{col(X,I)}.$$
$$\mathtt{used(I)} \leftarrow \mathtt{col(X,I)}.$$

$$\Leftarrow \mathtt{used(I)}.$$

The first rule guesses a coloring for the graph; the set of strong constraints checks the guess that two joined vertices do not have the same color, and that each vertex is assigned to exactly one color; the weak constraint requires that the number of colors used is minimum.

The query $\langle \mathtt{col(1,c1)}, \mathtt{MIN\_COLORING} \rangle$ is supplied on different graph topologies. Some of the results are shown in Figure 9.

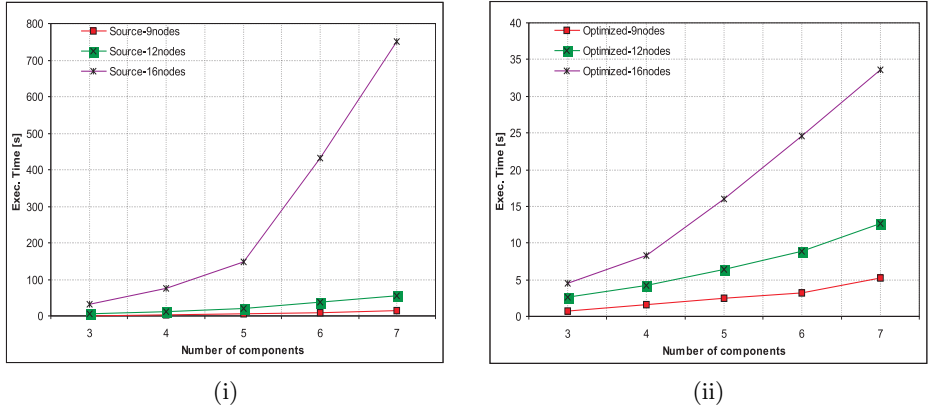

Fig. 9. Execution time for the Min coloring problem.

In particular the results obtained by augmenting the number of disconnected components, and by augmenting the cardinality of the components have been investigated. Figure 9.(i) and Figure 9.(ii) show, respectively, the performance of the optimized program and of the source program with a different number of components. In all the experiments the first component contains node 1.

## 6 Conclusions

In this paper a technique has been proposed for the optimization of bound queries over disjunctive deductive databases with constraints (a simple and expressive form of unstratified negation), which extends a previous technique suitable for disjunctive Datalog programs. As the usual way of expressing declaratively hard problems is based on the *guess-and-check* technique, where the *guess* part is expressed by means of disjunctive rules and the *check* part is expressed by means of constraints, the technique proposed here is highly relevant for the optimization of queries expressing hard problems.

The proposed approach is based on the use of a binding propagation technique which, by reducing the size of the data relevant to answer the query, is suitable for minimizing the complexity of computing a single model and the whole number of models to be considered.

The main contribution of the paper lies in the definition of a rewriting algorithm which systematically utilizes the query goal to propagate the binding through both the rules and the constraints thereby avoiding the computation of useless models. An interesting peculiarity of the formalization proposed here is that it is completely independent of the particular strategy adopted for propagating the binding: in this way, the results are completely orthogonal to the Magic-Set technique in itself, and, hence, to the results in (Greco, 2003). The value of the technique has been proved by several experiments.

*Acknowledgement.* The authors thank Nicola Leone and Wolfgang Faber for useful suggestions.